\documentclass[journal]{IEEEtran}

\renewcommand{\baselinestretch}{1}

\usepackage{epsfig,tabularx,amsfonts,amssymb,amsmath}
\usepackage{graphics} 
\usepackage[dvips]{color}
\usepackage{mathrsfs}
\usepackage{subfigure}
\usepackage{calc,ifthen}
\usepackage{multirow}
\usepackage{bm}
\usepackage{float}
\usepackage{pstricks}
\usepackage{amsthm}

\usepackage{empheq}
\usepackage{setspace}

\usepackage{cite}
\usepackage{psfrag}
\usepackage{pstool}
\usepackage{changes}
\usepackage{mathtools}

\usepackage{fancyhdr}

\usepackage{color}
\DeclareMathOperator{\sign}{sign}

\newcommand{\mynote}[1]{}
\newcommand{\newnote}[1]{}
\newcommand{\cancelled}[1]

\def\soc{{\rm SOC}}
\def\sign{\mathop{\rm sign}}

\def\emstwo {HPTS}

\title{Optimal Energy Management of Series Hybrid Electric Vehicles with 
Engine Start-Stop System}
\author{Boli Chen, Xiao Pan and Simos A. Evangelou
\thanks{B. Chen is with the Dept. of Electronic and Electrical Engineering at University College London, UK ({\tt\small
boli.chen@ucl.ac.uk}).}
\thanks{X. Pan is with the Dept. of Electrical and Electronic Engineering at Imperial College London, UK {\tt\small
(xiao.pan17@imperial.ac.uk)}.}
\thanks{S. A. Evangelou is with the Dept. of Electrical and Electronic Engineering at Imperial College London, UK
{\tt\small (s.evangelou@imperial.ac.uk)}.}
\thanks{This research was supported by the EPSRC Grant EP/N022262/1.}}

\begin{document}

\thispagestyle{empty}
\setcounter{page}{0}
\begin{figure*}
\centering
\includegraphics[width=.9\textwidth]{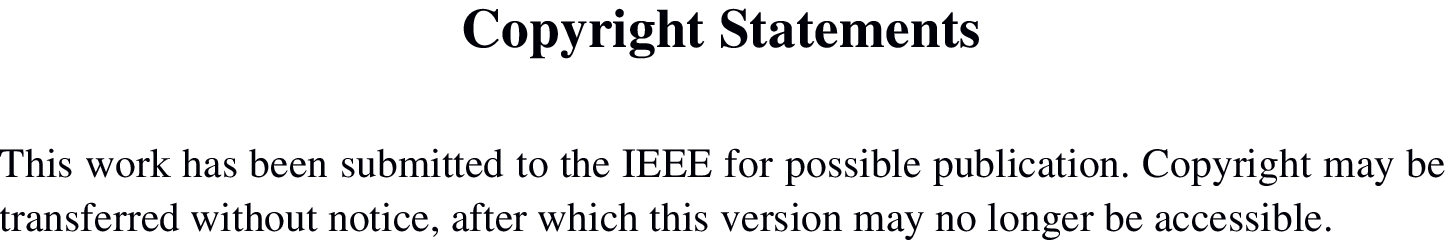}
\end{figure*}

\maketitle

\thispagestyle{fancy}
\chead{This work has been submitted to the IEEE for possible publication. Copyright may be transferred without notice, after which this version may no longer be accessible.}
\rhead{~\thepage~}
\renewcommand{\headrulewidth}{0pt}

\pagestyle{fancy}
\chead{This work has been submitted to the IEEE for possible publication. Copyright may be transferred without notice, after which this version may no longer be accessible.}
\rhead{~\thepage~}
\renewcommand{\headrulewidth}{0pt}



\begin{abstract}
This paper develops energy management (EM) control for series hybrid electric vehicles (HEVs) that include an engine start-stop system (SSS). The objective of the control is to optimally split the energy between the sources of the powertrain and achieve fuel consumption minimization. In contrast to existing works, a fuel penalty is used to characterize more realistically SSS engine restarts, to enable more realistic design and testing of control algorithms. 
The paper first derives two important analytic results: a) analytic EM optimal solutions of fundamental and commonly used series HEV frameworks, and b) proof of optimality of charge sustaining operation in series HEVs. It then proposes a novel heuristic control strategy, the hysteresis power threshold strategy ({\emstwo}), by amalgamating simple and effective control rules extracted from the suite of derived analytic EM optimal solutions. The decision parameters of the control strategy are small in number and freely tunable. The overall control performance can be fully optimized for different HEV parameters and driving cycles by a systematic tuning process, while also targeting charge sustaining operation. The performance of {\emstwo} is evaluated and benchmarked against existing methodologies, including dynamic programming (DP) and a recently proposed state-of-the-art heuristic strategy. The results show the effectiveness and robustness of the {\emstwo} and also indicate its potential to be used as the benchmark strategy for high fidelity HEV models, where DP is no longer applicable due to computational complexity.

\end{abstract}


\vspace{-4mm}
\section*{Abbreviations}
\noindent
\begin{tabular*}{0.49\textwidth}{@{}l@{\hspace{2mm}} @{\extracolsep{\fill}}l} 
 CS  & Charge Sustaining\\
 DP & Dynamic Programming\\
 EFC & Equivalent Fuel Consumption\\
 EM & Energy Management\\
 FCM & Fuel Consumption Model \\
 HEV & Hybrid Electric Vehicle\\
 {\emstwo} & Hysteresis Power Threshold Strategy\\
 ICE & Internal Combustion Engine\\
 PL & Propulsion Load\\
 PMP & Pontryagin's Minimum Principle\\
 PS & Primary Source of Energy\\
 SOC & State of Charge\\
 SS & Secondary Source of Energy\\
 SSS & Start-Stop System\\
 WLTP & Worldwide harmonized Light vehicles Test 
 Procedure\\
 
\end{tabular*}

\section{Introduction}
\IEEEPARstart{H}{ybrid} electric vehicles (HEVs) are regarded as an essential stage of transportation electrification towards addressing the global concerns 
of environmental pollution caused by emissions. In popular 
architectures the hybrid electric vehicle (HEV) 
embodies a combination of an internal combustion engine (ICE) and a battery-driven electric motor. As such, HEVs can profit from a freely optimized power split between the two energy sources for improving fuel economy as compared to 
conventional vehicles. Additionally, modern HEVs are usually equipped with an engine start-stop system (SSS), which automatically shuts down and restarts the ICE, thereby 
enabling a further reduction of idling fuel consumption and emissions.

The problem of finding a fuel-efficient power split for HEVs, known as the energy management (EM) control problem, has drawn considerable attention in the past decade. A comprehensive overview of existing EM techniques, from rule-based to optimization-based, can be found in \cite{Malikopoulos:2014,Wirasingha:2011,Clara:2017,Biswas:tvt2019}. Most of the existing EM approaches are optimization-based, such as dynamic programming (DP) \cite{koot:2005}, quadratic programming (QP) \cite{Chen:jps2014}, Pontryagin's Minimum Principle (PMP) \cite{Uebel:2018,chen:2019}, equivalent consumption minimization strategies (ECMS) \cite{serrao:2011,Musardo+Rizzoni+Staccia/cdc:2005}, extremum seeking \cite{Dincmen:vsd2012}, and model predictive control \cite{Ripaccioli/acc:2010,Yan:2012,Zhang:2016,zhang:2014}. More specifically, DP is able to guarantee global optimal solutions in general optimization problems, however, it is computationally intensive and therefore limited to highly simplified models \cite{Guzzella:TCST2007}. 
The equivalent consumption minimization strategy (ECMS) is derived using the Pontryagin's Minimum Principle (PMP), which results in a computational efficient local optimization algorithm at the expense of weakened optimality especially when the model complexity increases \cite{serrao:2011,Musardo+Rizzoni+Staccia/cdc:2005}. Model predictive control offers a computationally efficient alternative to global optimal control in yielding near-optimal solutions with less simple models. Moreover, stochastic and data-driven optimization (machine learning) methodologies are continually emerging and have become effective and important means to formulate model-free EM control strategies \cite{Hu:tiem2019,Liang:2017,liu:2017,murphey:2012,Murphey:2013}. For further improvement of the HEV fuel efficiency, engine on/off control needs to be embedded into EM control design to account for SSS dynamics, and therefore to avoid inefficient engine idling operation. The vast majority of the literature assumes the SSS to be ideal with no extra cost for engine restarts \cite{Uebel:2018,Roland:ecc18,wassif:2016,Markus:tvt2015}, which may lead to EM strategies that force the engine to a very rapid succession of starts/stops. To prevent this unnecessarily inefficient behaviour, \cite{Sciarretta+Back+Guzzella/IEEE:2004,Tobias:energies2014,Markus:2015} propose an enhanced SSS modelling framework in the context of conventional heuristic or numerical optimization of parallel HEV EM, where the fuel required to accelerate the engine from rest to idle speed is taken into account so that fast ICE start/stop transitions are penalized and avoided as a consequence.

Although optimization-based EM strategies can be easily tackled offline, they are usually not feasible for onboard computation units of modern HEVs due to the computational and memory limitations. However, due to the simplicity in implementation and ease of understanding of their operating principles, rule-based strategies are more prevalent among commercial HEVs. Rule-based EM strategies are usually characterized by Boolean or fuzzy rules, which are described as a set of rules that compute the control signals based on pre-established thresholds over the controlled variables \cite{Zhang:2011,wassif:2016,wassif:19}. The Thermostat Control Strategy (TCS) and Power Follower Control Strategy (PFCS) are the two most conventional rule-based techniques, yet their fuel economies are not optimized. The operational rules behind them are respectively load following and load leveling mechanisms that have been extensively used in rule-based EM techniques of HEVs. Nonetheless, the conventional TCS and PFCS are outdated for modern HEVs where the SSS are becoming ubiquitous and very efficient these days, allowing the engine to be turned off and on with very low fuel consumption penalty. Although the recently proposed exclusive operation strategy (XOS) \cite{wassif:2016} and optimal primary source strategy (OPSS) \cite{wassif:19} dramatically improve the optimality of the existing rule-based approaches, their performance still falls behind optimization-based benchmarks by some margin.

Previous work of the authors in \cite{boli:ecc19} has proposed an innovative rule-based control strategy for series HEVs that bridges the gap between rule-based and optimization-based methods. This strategy can emulate the globally optimal solution to the EM problem with simple and effective rules, which are extracted from the closed-form optimal EM solution for a simplified vehicle/powertrain model.
However, the investigation in \cite{boli:ecc19} is of limited scope since it does not consider the impact of the SSS and furthermore it provides an analysis, and consequently a control design, on the basis that battery charge sustaining (CS) operation (for a whole mission) is strictly guaranteed, without an investigation into the optimality (in terms of fuel economy) or otherwise of the CS operation.

The present paper works on bridging the gap between rule-based and optimization-based EM strategies in the more general context with the SSS considered, by proposing a novel heuristic strategy for EM control of series HEVs with SSS. The series HEV architecture, on which the paper focuses, is a common arrangement for modern HEVs \cite{xie:energy2019} and involves a number of products in the market, such as the Nissan Note e-power, VIA Motors products and numerous other extended-range electric vehicles. The main contributions of the work are summarized as follows: 
\renewcommand{\theenumi}{\alph{enumi}}
\begin{enumerate}
\item Fundamental analysis with an HEV model without engine SSS that considers the main physics of the EM problem for series HEVs is conducted and feasible fundamental solutions of the optimal EM for series HEVs are found (this is a generalization of the work in \cite{boli:ecc19}).\label{enu:noSSS}
\item Fundamental analysis with an HEV model with an ideal (lossless) engine SSS, which additionally to the model features in contribution \ref{enu:noSSS}) captures the basic physics of the SSS, is conducted and feasible fundamental solutions of the optimal EM for series HEVs with SSS are found.\label{enu:SSS}
\item Fundamental analysis proves that CS operation is a necessary condition to reach globally optimal fuel economy in series HEVs.\label{enu:CS}
\item By using simple and effective control rules that are inspired by the fundamental analysis and solutions of the optimal EM in contributions \ref{enu:noSSS}) and \ref{enu:SSS}), as well as by the CS operation optimality result in contribution \ref{enu:CS}), a novel heuristic strategy, the hysteresis power threshold strategy ({\emstwo}), is proposed for the EM control of series HEVs with a more realistic SSS for which engine restarts are associated with a fuel penalty. An analytic solution of the optimal EM is not feasible in this case.
\item The performance of the {\emstwo} is evaluated and benchmarked against DP solutions and a recently developed state-of-the-art rule-based method. Moreover, the influence of the penalty fuel for engine restarts is further investigated by simulation.
\end{enumerate}

The remainder of the paper is structured as follows. Section~\ref{sec:model} introduces the vehicle model and the formulation of the EM problem. A theoretical 
derivation of the optimal EM solutions is presented in Section~\ref{sec:fundamentalanalysis}, and from the analysis, the {\emstwo} is proposed. Section~\ref{sec:fundamentalanalysis} also includes the analysis that proves the optimality of CS operation. Simulation results and discussion are presented in Section~\ref{sec:simulation}. Finally, concluding remarks are given in Section~\ref{sec:conclusions}. 

\section{Modeling of the series hybrid electric vehicle and problem formulation}\label{sec:model}
The vehicle model is developed based on the series HEV system (without the engine SSS) described in \cite{chen:2019}. In particular the ICE and the electric machines are approximated by steady-state efficiency maps, while the power converter, inverter and transmission system are modeled by constant efficiency factors that take into account the possible energy losses. This vehicle model captures the essential physical characteristics of the powertrain components with a relatively low dynamic order, thereby being widely applicable for HEV analysis and control design \cite{zhou:2016,Jose:tvt2018,Serrao+Rizzoni/acc:2008}. In the following part of this section, the overall vehicle system is briefly introduced with particular emphasis on the modeling of the engine start-stop system, which has not been considered sufficiently in the past. 

\begin{figure}
  \centering
  \includegraphics[width=1\columnwidth]{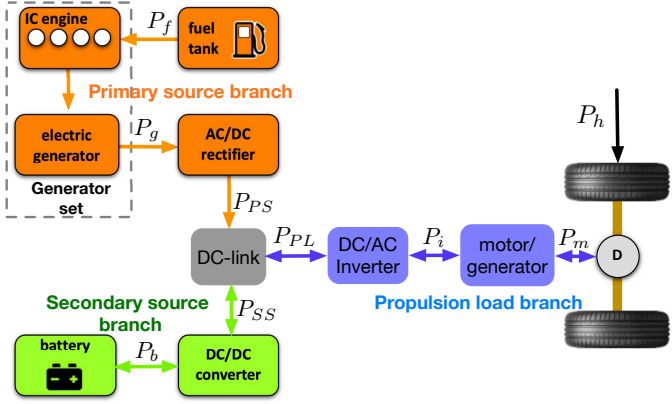}\\[-2ex]
  \caption{Powertrain architecture of the series HEV.}
  \label{fig:HEV-powertrain}
\end{figure}

\begin{figure}[ht!]
  \centering
  \includegraphics[width=0.95\columnwidth]{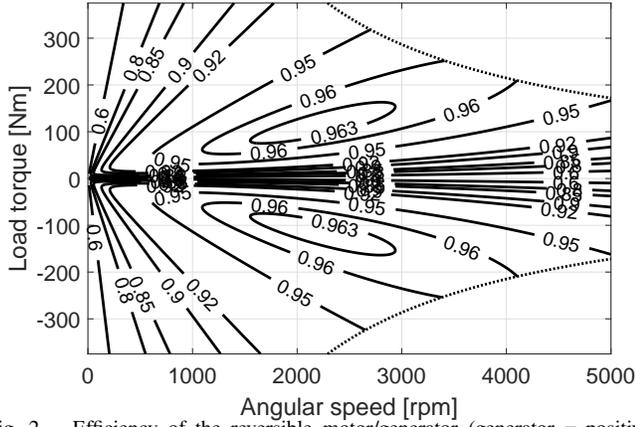}\\[-3ex]
  \caption{Efficiency of the reversible motor/generator (generator = positive torque, motor = negative torque) \cite{Evangelou+Shukla/ACC:2012}. The torque bounds (due to power limitation) are shown by dotted lines. The rated power of the machine is 95kW.}
  \label{fig:powertrain1}
\end{figure}

The series HEV powertrain architecture is sketched in Fig.~\ref{fig:HEV-powertrain}. 
As it can be noticed, the power outputs from the primary source (PS) and the secondary source (SS) branches are combined electrically at the DC-link. Then, the total power is delivered to the driving wheels included in the propulsion load (PL) branch, which is an inverter driven electric motor/generator, mechanically connected to the wheels with a transmission system characterized by a fixed drive ratio $\mathsf{g}_t$. The inverter and transmission are modeled as constant efficiency terms $\eta_i,\,\eta_t$, while the efficiency of the motor/generator, $\eta_m$, is described by a static efficiency map of the load torque and the angular speed, as shown in Fig.~ \ref{fig:powertrain1}. 

By virtue of the series architecture, it is reasonable to assume the load power $P_{PL}$ a known signal, as it is independent of the EM (power split) between the two energy sources \cite{chen:tvt2019}. Given a driving cycle and the steady state efficiencies $\eta_i,\,\eta_t$, $\eta_m$, $P_{PL}$ may be determined by:
\begin{subequations}  \label{eq:PPL}
    \begin{empheq}[left={P_{PL}=\empheqlbrace\,}]{align}
      &   \frac{P_{drive}}{\eta_i\eta_m \eta_t}\,,   \forall \, P_{drive} \geq 0 \label{eq:PPL1} \\
      &  (P_{drive}-P_h){\eta_i\eta_m \eta_t}\,,   \forall \, P_{drive} < 0\,,\label{eq:PPL2}
    \end{empheq}
\end{subequations}
where $P_{drive}$ is the power at the driving wheels requested to follow the driving cycle, and $P_h$ is the mechanical braking power directly applied to the wheels. Furthermore, $P_{drive}$ can be evaluated by Newton's laws of motion, as follows:
\begin{equation}
P_{drive} = v\left(m\,a + F_T + F_D +  mg\sin{\theta}\right),
\label{eq:drivingforce}
\end{equation}
where $v$, $a$ and $m$ are the speed, acceleration and mass of the vehicle respectively, $F_T = f_T\,m\,g$ and $F_D = f_D v^2$ are the resistance forces respectively due to tires and aerodynamics drag, and $\theta$ is the 
 road slope associated with the speed profile. By considering that the mechanical braking power is dissipated and $P_{SS_{\min}}$ is the maximum charging power of the SS, it is possible to always freely choose $P_h$ such that $P_{PL}$ in \eqref{eq:PPL2} for $P_{drive} < 0$ is as follows
\begin{equation}\label{eq:PPLneg}
 P_{PL}=\displaystyle \max \left({P_{drive}}{\eta_i\eta_m \eta_t}, \,P_{SS_{\min}}\right) \,,   \forall \, P_{drive} < 0\,,
\end{equation}
to maximize energy regeneration, and hence fuel economy. 

Consequently, it is reasonable to decouple the EM problem of a series HEV into two steps: 1) compute $P_{PL}$ requested by a driving cycle by \eqref{eq:PPL1}, \eqref{eq:drivingforce} and \eqref{eq:PPLneg}, and 2) find an appropriate power split (for $P_{PL}(t)>0$) between the two energy sources subject to the power balance at the DC-link (see \eqref{eq:powerbalance} described later in Section~\ref{subsec:model&problem}).

\subsubsection{Primary Source Branch} 
\begin{figure}[h!]
\centering
\includegraphics[width=0.49\columnwidth]{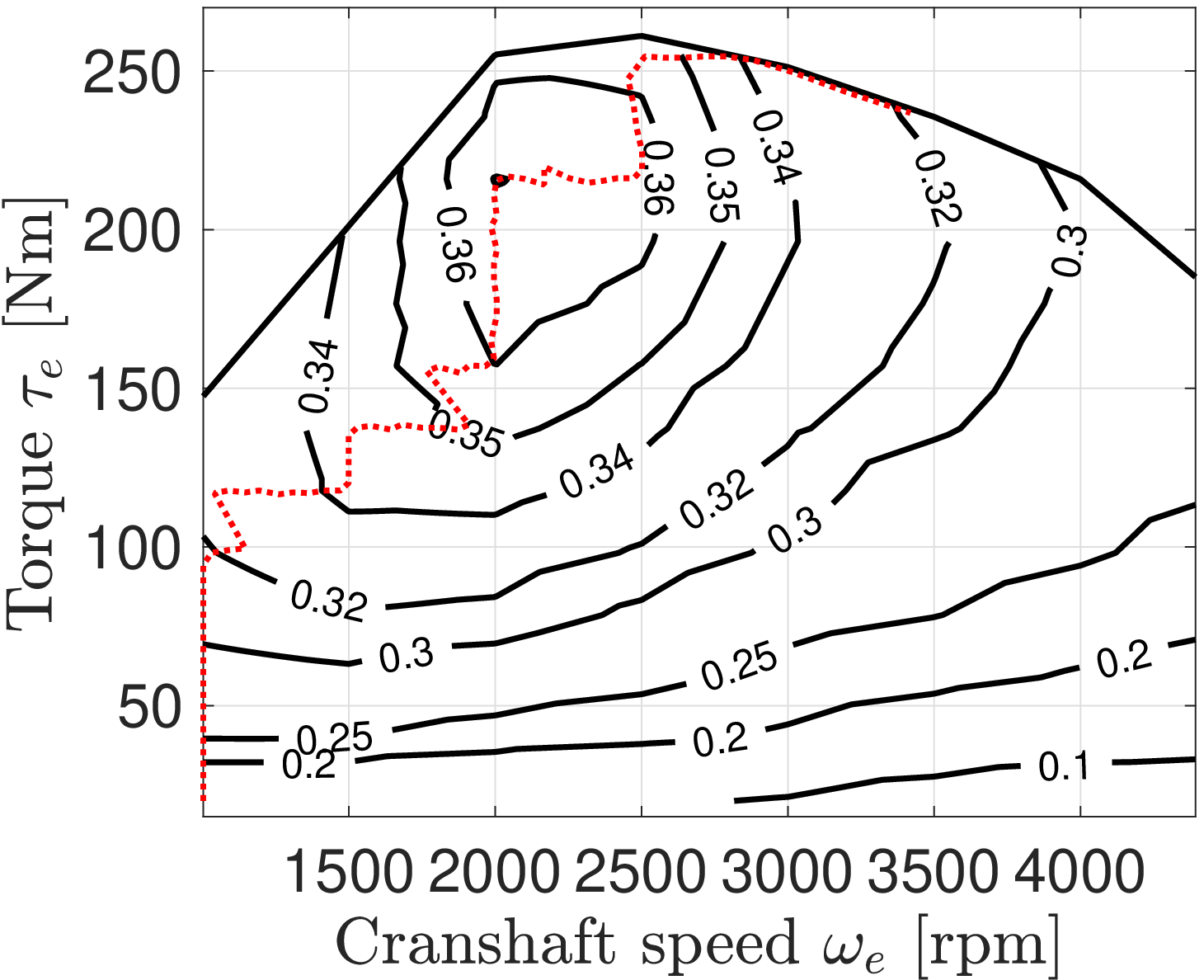}
\includegraphics[width=0.49\columnwidth]{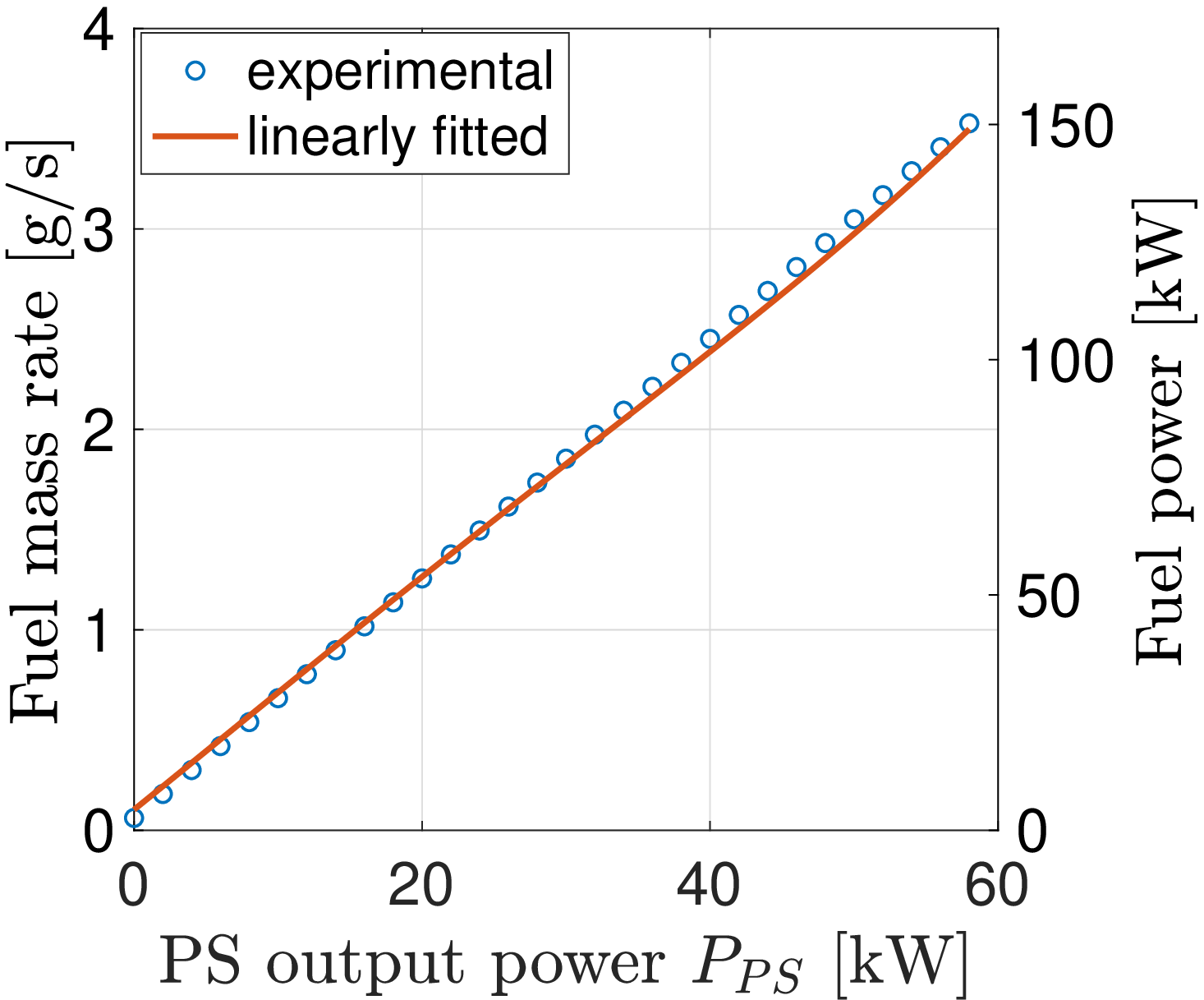}\\[-1.5ex] 
\caption{Left: map of overall efficiency of the engine branch. The torque-speed operating points for maximum engine branch efficiency at different output power values are shown by a dashed red curve. Right: fuel mass rate with PS power, when the most efficient torque-speed operating point is followed at each power value.}
\label{fig:fuelmassrate}
\end{figure}

As shown in Fig.~\ref{fig:HEV-powertrain}, the engine branch is formed by an ICE, a permanent magnet synchronous generator, and an AC-DC rectifier, which are connected in series. The overall efficiency of this branch is simply the product of individual component steady state efficiencies. The present work is based on the PS branch efficiency map shown in Fig.~\ref{fig:fuelmassrate} \cite{chen:tvt2019}.
%
The mechanical separation from the wheels allows the primary source branch to be constantly operated along the trajectory of the most efficient torque-speed operating points. In such a case, the fuel mass rate $q_f$, as shown in Fig.~\ref{fig:fuelmassrate}, can be fitted approximately as a linear function of the branch output power $P_{PS}$. The dynamic equation of the fuel mass $m_f$ is therefore given by,
\begin{equation}\label{eq:Q_f}
    \dot{m}_f = q_{f0} + \alpha_f P_{PS},
\end{equation}
in which $q_{f0}$ acts as the idling fuel mass rate, and $\alpha_f$ is the coefficient of power transformation.

Furthermore, the start-stop system allows the ICE to be switched off without idling loss. However, some amount of fuel is consumed to turn on the ICE again. In this work, this penalty fuel usage is modeled by a constant term $m_{p} = Kq_{f0}\,$kg, which is equivalent to the amount of fuel consumed by idling the ICE for $K$ seconds. Moreover, the delay of restarting the engine is neglected as it mainly affects the driving comfort rather the than fuel economy. 
To integrate the SSS dynamics into the fuel consumption model (FCM) \eqref{eq:Q_f}, let us introduce the binary engine off/on state $s \in \{0, 1\}$ and the jump set $\mathcal{S} \triangleq \{s|\,s^+ \ne s\}$, with $s^+$ the next value of the state. 
As such, ICE operation can be characterized in terms of $s$ and $P_{PS}$: 1) the engine is switched off when $s = 0, \,P_{PS} = 0$, 2) the engine is idling when $s = 1, \,P_{PS} = 0$, and 3) the engine produces propulsive power when $s = 1, \,P_{PS} >0$. Finally, the FCM in the presence of the SSS may be described by the following hybrid dynamical system: 
\begin{equation}\label{eq:hybridsys1}
\left\{
\begin{array}{lll}
    \dot{m}_f  = q_{f0}\,s + \alpha_f P_{PS}  & \text{if } s \notin \mathcal{S}  \\
     \dot{s}  = 0
\end{array}\right.
\end{equation}
and 
\begin{equation}\label{eq:hybridsys2}
\left\{
\begin{array}{lll}
     s^+  =  s+u_{s}  & \text{if } s \in \mathcal{S}\\
     m_f^+  = m_f + m_{p}(1-s)  
\end{array}\right.
\end{equation}
where $u_s \in \{-1,\,1\}$ is the SSS control signal.

\subsubsection{Secondary Source Branch}
The battery is modeled as a series connection of an ideal voltage source and an ohmic resistance, therefore the battery voltage $V_b$ is defined by 
$V_b = V_{oc} - i_b  R_b$, where $V_{oc}$ is the open circuit voltage of the battery, $i_b$ is the battery current assumed positive during the discharge phase, and $R_b$ is the internal resistance. By considering the battery power $P_b =V_b i_b$, $i_b$ can be solved with respect to $V_{oc}$, $R_b$ and $P_b$, as follows:
\begin{equation}
\label{eq:i_b}
i_b = \frac{V_{oc}-\sqrt{V^2_{oc}-4P_b  R_b}}{2 R_b}\,.
\end{equation}
The battery state-of-charge (SOC) represents the only state variable, governed by $\dot{\soc} = -{i_b}/{Q_{\max}}$, where $Q_{\max}$ is the battery capacity.  
Instead of using a nonlinear mapping between SOC and the open circuit voltage, $V_{oc}$ is reasonably approximated by a constant voltage, which is compatible with the usual aim of a charge sustaining (CS) battery management, by which the SOC is narrowly constrained. 
Furthermore, by combining the battery with the bidirectional DC/DC converter, the SS output power is obtained by:
\begin{equation}\label{eq:eta_dc}
  P_{SS} = \eta_{dc} ^{\sign(P_{SS})}  P_b,
\end{equation}
where $\eta_{dc}$ is the efficiency of the DC/DC converter.
Substituting the algebraic solution of $i_b$ (obtained by applying \eqref{eq:eta_dc} in \eqref{eq:i_b}), the dynamic behavior of the SS can be described by the differential equation of SOC with respect to $P_{SS}$ only as:
 \begin{equation}
\dot{\soc} =  \frac{-V_{oc}+\sqrt{V_{oc}^2 -\displaystyle {4 P_{SS}\,R_b} /{\eta_{dc} ^{\sign(P_{SS})}}}}{2 R_b\,Q_{\max}}\,.
\label{eq:dsoc}
\end{equation}

\subsubsection{Overall model and energy management problem} \label{subsec:model&problem}
In view of \eqref{eq:hybridsys1}, \eqref{eq:hybridsys2} and \eqref{eq:dsoc}, the overall system dynamics are captured by the hybrid system given by: 
\begin{equation}\label{eq:simplemodel1}
     \dot{\textbf{x}} = \mathbf{f}(\textbf{x},\textbf{u})
    \!=\!
    \begin{pmatrix}
    \displaystyle {q}_{f0}s + \alpha_f P_{PS}  \\
    \displaystyle \frac{-V_{oc}+\sqrt{V_{oc}^2 -\displaystyle {4 P_{SS}\,R_b} /{\eta_{dc} ^{\sign(P_{SS})}}}}{2 R_b\,Q_{\max}}\\0  
    \end{pmatrix},
\end{equation}
if $s \notin \mathcal{S}$, and if $s \in \mathcal{S}$:
\begin{equation}\label{eq:simplemodel2}
     \textbf{x}^+ = \mathbf{g}(\textbf{x},\textbf{u})
    \!=\!
    \begin{pmatrix}
    m_f+m_{p}(1-s)  \\
    \soc\\
    s+u_s  
    \end{pmatrix},
\end{equation}
where  ${\bf x} = [m_f,\,\soc,\,s]^{\top}$ and ${\bf u} \triangleq [P_{PS},\,P_{SS},\,u_s]^{\top}$ represent the state variables and control inputs, respectively. 

The EM control aims to minimize the overall fuel consumption $m_f$ by an appropriate power split between $P_{PS}$ and $P_{SS}$, which satisfy the DC-link power balance:
\begin{equation}
P_{PL}= s\,P_{PS}+P_{SS}\,.
\label{eq:powerbalance}
\end{equation}
Moreover, the operation of both energy sources is subject to,
\begin{align}
\soc_{\min} \leq \soc \leq \soc_{\max}\,, \label{eq:OCPconstraint1} \\
0 \leq P_{PS} \leq P_{PS_{\max}}\,, \label{eq:OCPconstraint2} \\
P_{SS_{\min}} \leq P_{SS} \leq P_{SS_{\max}}, \label{eq:OCPconstraint3}
\end{align}
where $\soc_{\min}$ and $\soc_{\max}$ are the SOC operational limits, and $P_{PS_{\max}}$ and $P_{SS_{\max}}$ are the maximum propulsive power PS and SS can deliver respectively. The main characteristic parameters of the vehicle model are summarized in Table~\ref{tab:vehicle data}, where the power limits are chosen to emulate the energy sources for a non-plug-in HEV.
\begin{table}[!ht]
\centering
\caption{Main Vehicle Model Parameters}\vspace{-2mm}
\label{tab:vehicle data}
\begin{tabular*}{1\columnwidth}{l @{\extracolsep{\fill}} cl}
\hline
\hline
 symbol & value & description\\
 \hline
 $m$ & 1500~kg & vehicle mass \\
 $f_T$ & $0.01\,$ & rolling resistance coefficient\\
 $f_D$ & $0.47$ & aerodynamics drag coefficient \\
 $\eta_{t}$ & $0.96$ & efficiency of the transmission \\
 $\mathsf{g}_t$ & $10$ & transmission ratio\\
 $q_{f0}$ & $0.12$g/s & idling fuel mass rate\\
 $\alpha_f$ & $0.059$ g/kW/s & power transformation factor\\
 $Q_{\max}$ & 5~Ah & battery capacity \\
 $R_{b}$ & $0.2056$~$\Omega$ & battery internal resistance \\
 $V_{oc}$ & $300\,$V & battery open circuit voltage \\
 $\eta_r,\,\eta_{i}$ & $0.96$ & efficiency of inverters\\
 $\eta_{dc}$ & $0.96$ & efficiency of converter \\
 $P_{SS_{\min\!/\!\!\max}}$    &$-15/30$kW & SS power limits\\
 $\soc_{\min\!/\!\!\max}$    &$0.5/0.8$ & battery SOC limits\\
 $P_{PS_{\max}}$    &$70$kW & PS power limit\\
\hline
\hline
\end{tabular*}
\end{table}

\section{Energy management strategies based on the fundamental analysis of fuel efficiency optimization}\label{sec:fundamentalanalysis}
This section aims to provide a fundamental analysis to show the nature of optimal EM solutions using two variants of the presented vehicle model, both of which have been used for the design of EM strategies in the literature: 1) without engine SSS, and 2) with an ideal engine SSS ($m_{p} = 0$). The analytic solutions yield some fundamental principles that are used to go beyond the treatment of models 1) and 2) and construct a new heuristic control strategy, the {\emstwo}, for the more realistic model \eqref{eq:simplemodel1}-\eqref{eq:simplemodel2}. Further analysis is carried out to justify the optimality of the CS operation in terms of fuel efficiency, which is linked to the control design and results presented in this paper.

For the sake of further analysis, let us first introduce some useful notations and definitions for the upcoming analysis. Consider $T$ the total time of the driving mission. Two sets of time intervals $\Phi \triangleq \{t| P_{PL}(t)\geq 0\}$ and $\Psi \triangleq \{t|P_{PL}(t) <0\}$ are considered, such that $\Phi \cup \Psi$ is the full time horizon $\{t| 0\leq t \leq T\}$. 
The overall SOC variation over $[0,T]$ is defined as
$
\Delta \text{\soc} \triangleq \soc(0) - \soc(T).
$ 
It is clear that, 
\[
\Delta \text{\soc} = \Delta_{\Phi} \text{\soc}+\Delta_{\Psi} \text{\soc},
\] 
where, 
$
  \Delta_{\Phi} \soc \triangleq -\!\int_{\Phi}\frac{d\soc}{dt}dt\,, 
\Delta_{\Psi} \soc  \triangleq -\!\int_{\Psi}\frac{d\soc}{dt}dt \leq 0.
$
Moreover, $\Phi$ can be divided into 2 subsets as $\Phi= \Phi_d \cup \Phi_c$ with
$\Phi_d \triangleq \{t|P_{PL}(t)\geq 0, i_b(t) \geq 0\}$, $\Phi_c\triangleq \{t| P_{PL}(t)\geq 0, i_b(t) < 0\}.$
These subsets respectively collect the battery discharging and charging intervals for all $t\in \Phi$. Therefore,
\begin{equation}\label{eq:DeltaSOC}
\Delta \text{\soc} = \Delta_{\Phi_d} \soc+\Delta_{\Phi_c} \soc+\Delta_{\Psi} \text{\soc}
\end{equation}
with $\Delta_{\Phi_c} \soc<0 $ and $\Delta_{\Phi_d} \soc \geq 0$.
Finally, for brevity, the dependence of all variables on $t$ is dropped in the following analysis.

\subsection{Analysis for the vehicle model without the SSS}\label{subsec:fundamental1}
Let us start from analyzing the powertrain system without the SSS, as the analytic solution is also instrumental for subsequently studying the optimal EM solution for the system with consideration of the SSS. 

According to the power balance at the DC-link \eqref{eq:powerbalance}, the input power signal $P_{PS}$ of the FCM \eqref{eq:Q_f} can be replaced by $P_{PS} = P_{PL}-P_{SS}$ ($s$ is substituted as 1). By considering \eqref{eq:Q_f}, the fuel consumption minimization problem,  $\min J = m_f(T)$, 
can be rewritten as, 
\begin{multline*}
\min J = \int_0^T q_{f0}\,dt+\alpha_f \int_0^T P_{PL} \,dt -\alpha_f \int_0^T P_{SS}\,dt,
\end{multline*}
which is equivalent to
\begin{equation}\label{eq:objective}
\min_{P_{SS}} J = -\alpha_f \int_0^T P_{SS}\,dt,
\end{equation}
as $\int_0^T q_{f0}\,dt+\alpha_f \int_0^T P_{PL} \,dt$ is constant for a given driving cycle and independent of the EM control. The EM control problem of a series HEV is now formulated as an optimization problem with only one dynamic state, $\soc$, and a single control input, $P_{SS}$, subject to the SOC operational limits \eqref{eq:OCPconstraint1} and the energy source power limits,
\begin{equation}\label{eq:OCPconstraint5}
\max(P_{PL}\!-\!P_{PS_{\max}},\,P_{SS_{\min}}) \leq P_{SS} \leq \min(P_{SS_{\max}},P_{PL}),
\end{equation}
where \eqref{eq:OCPconstraint5} is obtained by combining \eqref{eq:powerbalance}, \eqref{eq:OCPconstraint2} and \eqref{eq:OCPconstraint3}. Based on the optimal solution for $P_{SS}$, the fuel usage, $m_f(T)$, can be evaluated a posteriori. 
To address this optimization problem, constrained and unconstrained arcs need to be pieced together. Based on the state constraints \eqref{eq:OCPconstraint1}, the optimal solution is the result of different combinations of the following possible arcs.
\subsubsection{State Constraints not Active}


According to the PMP, a candidate for an optimal control input $P_{SS}^*$ for minimizing \eqref{eq:objective} is found if $P_{SS}^*$ minimizes the Hamiltonian:
\begin{equation}
H \!=\!-\alpha_f P_{SS}+\left\{
\begin{array}{lr}
 \!\!\!\!\lambda \displaystyle \frac{-V_{oc}\!+\!\sqrt{V_{oc}^2 \!-\!\displaystyle {4 P_{SS}\,R_b} /{\eta_{dc}}}}{2 R_b\,Q_{\max}},&\!\!\!\! P_{SS}\geq 0\\
\!\!\!\! \lambda \displaystyle\frac{-V_{oc}\!+\!\sqrt{V_{oc}^2 \!-\!\displaystyle {4 P_{SS}\,R_b \eta_{dc}}}}{2 R_b\,Q_{\max}},&\!\!\!\! P_{SS} < 0
\end{array}\right.
\label{eq:hamiltonian}
\end{equation}
which includes the costate $\lambda$. The dynamics of the costate are described by
\begin{equation}
\dot{\lambda} = -\frac{\partial H}{\partial \soc} = 0,
\label{eq:costatedyna}
\end{equation}
which implies the optimal costate $\lambda$ is constant. By taking the partial derivative of $H$ with respect to $P_{SS}$, we obtain:
\[
\frac{\partial H}{\partial P_{SS}} = -\alpha_f-\left\{
\begin{array}{lr}
\!\!\displaystyle \frac{\lambda}{\eta_{dc} Q_{\max}\sqrt{V_{oc}^2 -\displaystyle {4 P_{SS}\,R_b}/\eta_{dc}}},& P_{SS}\geq 0\\
\!\! \displaystyle \frac{\lambda \eta_{dc}}{Q_{\max}\sqrt{V_{oc}^2 -\displaystyle {4 P_{SS}\,R_b}\eta_{dc}}},& P_{SS} < 0
\end{array}\right.
\]
If $\lambda = 0$, ${\partial H}/{\partial P_{SS}} = -\alpha_f<0$, which is independent of the input. Hence, the Hamiltonian is minimized at the maximum value of the input, as follows:
\begin{equation}
P_{SS}^* = P_{SS_{\max}}.
\label{eq:uopt1}
\end{equation}
If $\lambda > 0$, it is immediate to show that 
${\partial H}/{\partial P_{SS}} < 0,\ \forall P_{SS}\,$ (with $P_{SS_{\max}}<V_{oc}^2\eta_{dc}/(4R_b)$), 
and the optimal control input follows \eqref{eq:uopt1}. 
If $\lambda < 0$, the second order derivative of $H$ with respect to $P_{SS}$, 
\[
\frac{\partial^2 H}{\partial P_{SS}^2} = \left\{
\begin{array}{lr}
\!\!\displaystyle -\frac{2\lambda R_b}{\eta_{dc}^2 Q_{\max}\left({V_{oc}^2 -\displaystyle {4 P_{SS}\,R_b}/\eta_{dc}}\right)^{3/2}},& P_{SS}\geq 0\\
\!\! \displaystyle -\frac{2 \lambda \eta_{dc}^2 R_b}{Q_{\max}\left({V_{oc}^2 -\displaystyle {4 P_{SS}\,R_b}\eta_{dc}}\right)^{3/2}},& P_{SS} < 0
\end{array}\right.
\]
is always positive.  As such, $H$ is formed by two convex segments continuous at $H(0) = 0$, and the global minimum of $H$ depends on the minima of the functions representing both segments of $H$. By solving the algebraic equation ${\partial H}/{\partial P_{SS}} = 0$ for $\,P_{SS} \geq 0$ and $P_{SS}<0$ respectively, we obtain, 
\begin{subequations}
\begin{align}
&P_{SS}^* =\frac{1}{4 R_b} \left( \eta_{dc}V_{oc}^2-\frac{\lambda^2}{\alpha_f^2 Q_{\max}^2 \eta_{dc}} \right), \,\,\text{if}\,\, P_{SS} \geq 0,
\label{eq:optimalpss1}\\
&P_{SS}^* =\frac{1}{4 R_b} \left( \frac{V_{oc}^2}{\eta_{dc}}-\frac{\lambda^2 \eta_{dc}}{\alpha_f^2 Q_{\max}^2 } \right),\,\,\text{if}\,\,P_{SS}<0, \label{eq:optimalpss2}
\end{align}
\end{subequations}
at which the two quadratic functions reach their minima. If $\lambda \in (-\alpha_f Q_{\max} \eta_{dc} V_{oc},0)$, both optimal control inputs shown in \eqref{eq:optimalpss1} and \eqref{eq:optimalpss2} are positive, which implies the minimum of $H$ within the domain $P_{SS} < 0$ is $H(0)$, and therefore, the global minimum of $H$ is obtained at \eqref{eq:optimalpss1}.   
Similarly, for $\lambda \in (-\infty, -{\alpha_f Q_{\max} V_{oc}}/{\eta_{dc}})$, it can be inferred that both solutions shown in \eqref{eq:optimalpss1} and \eqref{eq:optimalpss2} are negative. Hence, the global minimum of $H$ is obtained when $P_{SS}$ follows \eqref{eq:optimalpss2}.   
Finally, if $\lambda \in \left[-{\alpha_f Q_{\max} V_{oc}}/{\eta_{dc}},-
\alpha_f Q_{\max} \eta_{dc} V_{oc}\right]$, the optimal control shown in \eqref{eq:optimalpss1} is negative and in \eqref{eq:optimalpss2} is positive. As such, $H$ is monotonically increasing for $P_{SS} \geq 0$ and monotonically decreasing for $P_{SS} < 0$, which yields $P_{SS}^* = 0$.
By taking into account the control constraints \eqref{eq:OCPconstraint5}, the optimal input $P_{SS}^*$ is given as follows:
\begin{equation}\label{eq:uoptphi}
P_{SS}^* \!\!= \!\! \left\{
\begin{array}{lll}
\!\!\!\! \min(P_{SS_{\max}},P_{PL}),\hspace{2.8cm} \forall \lambda \in [0,\infty),\\
\!\!\!\! \min\left( \! P_{SS}^{*+},P_{SS_{\max}},P_{PL} \right)\!\!,   \forall \lambda \!\in\! (-
\alpha_f Q_{\max} \eta_{dc} V_{oc},0),\\
\!\!\!\! \min(0,P_{PL}), \, \! \forall \lambda \!\in\! \left[\displaystyle -\frac{\alpha_f Q_{\max} V_{oc}}{\eta_{dc}},-
\alpha_f Q_{\max} \eta_{dc} V_{oc}\right]\!,\\
\!\!\!\! \min\left(\max\left( \! P_{SS}^{*-}\!,P_{PL}\!\!-\!\!P_{PS_{\max}},P_{SS_{\min}} \! \right),P_{PL}\right)\!\!, \\ \hspace{3.1cm} \forall \lambda \in \left(\!-\infty, \displaystyle -\frac{\alpha_f Q_{\max} V_{oc}}{\eta_{dc}}\!\right).
\end{array}\right.
\end{equation}
where, for brevity, we denote $P_{SS}^{*+}$ and $P_{SS}^{*-}$ the solutions given in \eqref{eq:optimalpss1} and \eqref{eq:optimalpss2}, respectively.
As it can be seen, there exist four possible optimal modes of operation depending on the value of $\lambda$. The numerical calculation of the closed form control solution for a given $P_{PL}$ profile involves identifying the constant operating power of the SS, which is equivalent of finding the costate $\lambda$. Given an $\soc(0)$ in conjunction with $P_{PL}$, $\lambda$ can be identified by a simple parameter searching approach that ends when the $\soc(T)$ is fulfilled. It is noteworthy that $P_{SS}^*$ is constant unless a control constraint (see \eqref{eq:OCPconstraint5}) is reached. Moreover, when $P_{PL}<0$ (i.e., $t\in \Psi$), it can be seen in all cases shown in \eqref{eq:uoptphi} that the optimal input is simply expressed as,
\begin{equation}
P_{SS}^* = P_{PL},
\label{eq:uoptneg}
\end{equation}
unless $\max\left( P_{SS}^{*-},P_{PL}-P_{PS_{\max}},P_{SS_{\min}}\right)<P_{PL}$ (which is a special case of the fourth case in \eqref{eq:uoptphi}), and in such a case, 
   \begin{equation}
P_{SS}^* = \max\left( P_{SS}^{*-},P_{PL}-P_{PS_{\max}},P_{SS_{\min}}\right).
\label{eq:uoptneg2}
\end{equation}
The latter is a non typical case where significant battery charging is required, such that the ICE will be active even during the braking phase to boost the battery charging power.

%

\subsubsection{State Constraints Active}\label{para:activeconstaint} 
By operating the SS at $P_{SS}^*$, the unconstrained optimal state trajectory, $\soc^*(t, P_{SS}^*)$, may violate the state constant \eqref{eq:OCPconstraint1} during the operation. The optimal solution in such cases can be found by invoking a recursive scheme \cite{Thijs:automatica2014}. Suppose that at some time $t=t_p$, the state constraint is exceeded the most in the unconstrained optimal trajectory, the problem is then split in two subproblems with boundary conditions $\{\soc(0), \soc(t_p)\}$ and $\{\soc(t_p),\soc(T)\}$, respectively, with $\soc(t_p)=\soc_{\max}$ in the case the upper state constraint is exceeded, otherwise $\soc(t_p)=\soc_{\min}$. By following the same approach used for the unconstraint case, it is immediate to find the optimal costate and the associated control solution for both problems. From the jump conditions of the PMP, $\lambda$ is discontinuous at $t_p$, and $\lambda(t_p^+)<\lambda(t_p^-)$ if the upper bound is reached, $\lambda(t_p^+)>\lambda(t_p^-)$ if the lower bound is reached. Once the solution for a subproblem is found, such properties can be utilized to facilitate the searching of $\lambda$ for the other subproblem. If the constraint is still violated in any of the two subproblems, the procedure is repeated until all state constraints are met. 

\begin{figure}[htp!]
  \centering
  \includegraphics[width=\columnwidth]{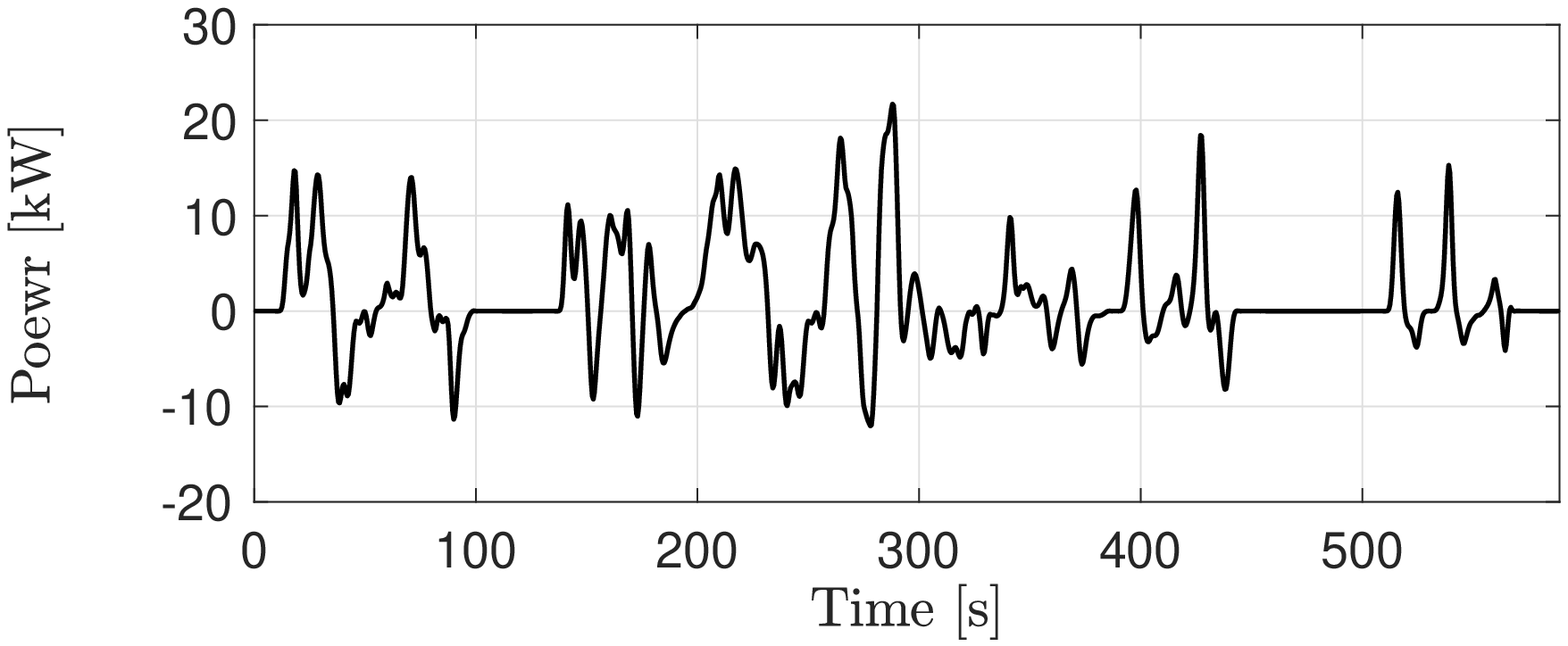}\\[-4.3ex]
    \includegraphics[width=\columnwidth]{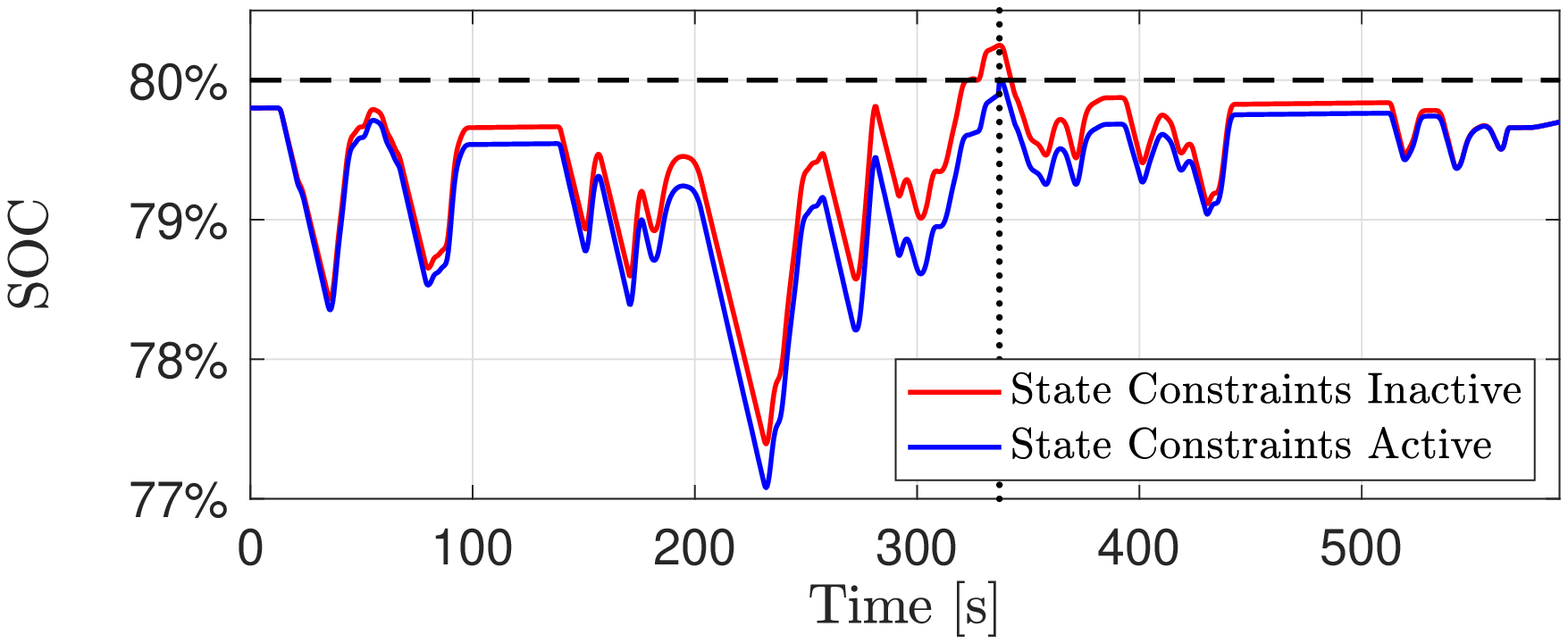}\\[-4.3ex]
    \includegraphics[width=\columnwidth]{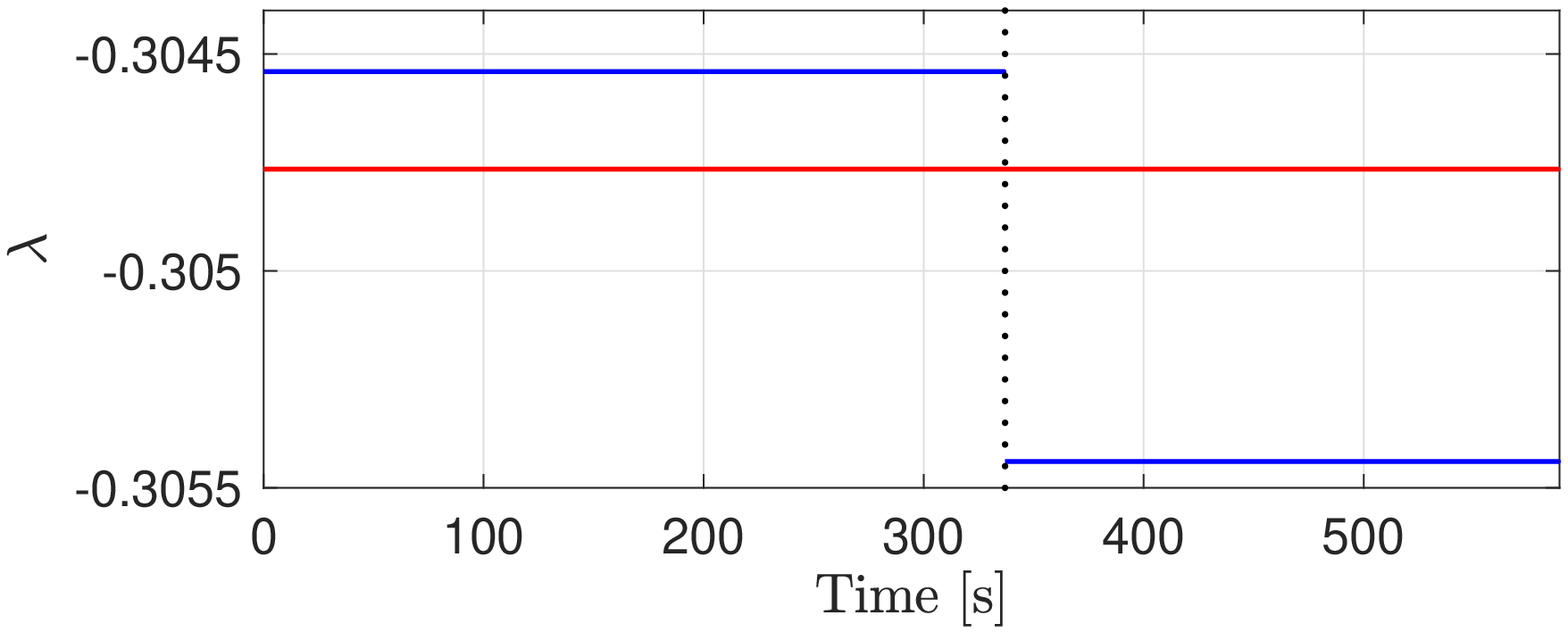}\\[-2ex]
  \caption{Sketch of example optimal solution for a vehicle mission when a state (SOC) constraint is reached. Top: Mission shown as $P_{PL}$ power profile results from the WL-L driving cycle. Middle: Optimal SOC trajectories for the unconstrained (red) and constrained (blue) cases. Bottom: Optimal costate, $\lambda$, for the unconstrained (red) and constrained (blue) cases.}
  \label{fig:discontinued_example}
\end{figure}

To illustrate the above recursive solution searching mechanism, a numerical example with HEV parameters given in Table~\ref{tab:vehicle data} is demonstrated in Fig.~\ref{fig:discontinued_example}. 
The HEV is requested to follow a segment (WL-L) of a standard driving cycle (that will be properly introduced later in Fig.~\ref{fig:WLTP}) with the associated $P_{PL}$ profile as shown in Fig.~\ref{fig:discontinued_example}, and the boundary conditions of SOC set to 
$\soc(0)=79.8\%, \soc(T)=79.8\%$. 
As it can be noticed, in the unconstrained optimal state (SOC) trajectory, the SOC constraint is exceeded the most at $t=337$s, where the boundary value problem is subsequently split into two subproblems. By repeating the global search of $\lambda$ for both subproblems, a piecewise constant $\lambda$ is found. Since the SOC constraint is fulfilled by the resulting state trajectories in both phases, the recursive algorithm ends and the optimal solution is found.

%

\begin{figure}[htb!]
\centering
\includegraphics[width=.95\columnwidth]{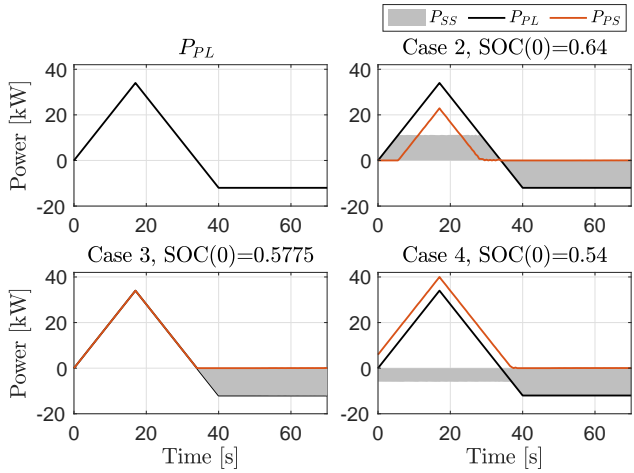}\\[-2ex]
\caption{Optimal power-split solutions found by DP for the case without SSS for an example power demand trajectory $P_{PL}$ of 70\,s, an $\soc(T)=0.65$ and three different cases of $\soc(0)$. Each $\soc(0)$ case is chosen to respectively actualize the last three of the closed-form solution cases in \eqref{eq:uoptphi}, associated with $\lambda=-0.2929$ for Case 2, $\lambda=-0.3448$ for Case 4 and  
$\lambda \in [-0.3361,-0.3097]$ for Case 3 (the control signal does not depend on $\lambda$), with the two $\lambda$ transition points being $-{\alpha_f Q_{\max} V_{oc}}/{\eta_{dc}}\!=\!-0.3361$ and $-\alpha_f Q_{\max} \eta_{dc} V_{oc}\!=\!-0.3097$. Note that in Case 4 for an even lower $\soc(0)$ (more demand on battery charging), during negative $P_{PL}$ the optimal solution may request the engine to contribute to the charging of the battery, and therefore $P_{PS}$ may become positive, instead of $0$, while at the same time $P_{SS}$ may become more negative than $P_{PL}$, instead of being equal to $P_{PL}$.}
\label{fig:example_withoutSSS}
\end{figure}

 To further clarify the closed form control solution \eqref{eq:uoptphi}, another example is carried out by utilizing a simple but representative $P_{PL}$ profile. As it can be seen in Fig.~\ref{fig:example_withoutSSS}, by following a simple $P_{PL}$ profile the optimal power-split profiles are found by DP for vehicle parameters given in Table~\ref{tab:vehicle data} with a fixed terminal $\soc(T)$ at $0.65$. To emulate different scenarios in \eqref{eq:uoptphi}, (last three Cases in \eqref{eq:uoptphi} except Case 1 which is a sub-solution of Case 2) initial SOC cases are set to: $0.64,\,0.5775,\,0.54$, which are associated with the three different cases of $\Delta_{\Phi} \soc>0$ (solution Case 2), $\Delta_{\Phi} \soc = 0$ (solution Case 3) and $\Delta_{\Phi} \soc < 0$ (solution Case 4), respectively. The subcase of Case 4 shown by \eqref{eq:uoptneg2} may occur if $\soc(0)$ is set to a further lower value (more charge is required to meet the terminal SOC condition by the end of the mission). Under these circumstances, the SS is charged throughout the mission at a constant power below the negative $P_{PL}$ value (this subcase is not shown in Fig.\ref{fig:example_withoutSSS}). The numerical (DP) results in Fig.~\ref{fig:example_withoutSSS} verify the closed form solution \eqref{eq:uoptphi}, indicating that for an optimal EM (during the propulsive phase $\Phi$) the SS is operated at a load leveling fashion unless a control or a state limit is reached (for example, during the period $t\in [0,5]$ in Case 2). As a consequence, once the PS is active, its power output $P_{PS}$ follows the trends of the $P_{PL}$ profile but with a fixed power difference that is equal to $P_{SS}^*$. 

%
\subsection{Analysis for the vehicle model with the lossless SSS}
\label{subsec:fundamental2} 
In this section, the SSS is integrated with no penalty fuel ($m_{p}\!=\!0$), as commonly 
assumed in EM studies (see for example \cite{Jose:tvt2018,Uebel:2018,Iman:tvt2021}).
In such a case, the new Hamiltonian $\forall t$ is:
\begin{equation}
H \!=\! \left\{
\begin{array}{lll}
\!\!\!H_1 = q_{f0}+\alpha_f(P_{PL}-P_{SS})\\ \hspace{1mm} +\lambda \displaystyle \frac{-V_{oc} +\sqrt{V_{oc}^2 \!-\!\displaystyle {4 P_{SS}\,R_b} /{\eta_{dc}^{\sign(P_{SS})}}}}{2 R_b\,Q_{\max}},  P_{SS}<P_{PL}\\
\!\!\!H_2 = \lambda \displaystyle \frac{-V_{oc}\!+\!\sqrt{V_{oc}^2 \!-\!\displaystyle {4 P_{PL}\,R_b} /{\eta_{dc}^{\sign(P_{PL})}}}}{2 R_b\,Q_{\max}}, \\ \hspace{6cm} P_{SS}=P_{PL}
\end{array}\right.
\label{eq:hamiltonian2}
\end{equation}
The dynamics of the costate remains the same as \eqref{eq:costatedyna}. The minimum of the Hamiltonian with respect to the control input $P_{SS}$ can be found by identifying the minimum of the two candidates $H_1$ and $H_2$ of the piecewise Hamiltonian \eqref{eq:hamiltonian2}, and 
 selecting the minimum between the two candidates \cite{Markus:2015,Iman:tvt2021}. For the sake of further analysis, let $H_1^*$ and $H_2^*$ denote the optima of $H_1$ and $H_2$, respectively. The minimum for the second candidate $H_2$ is trivial, 
\begin{equation}
H_2^* =H_2,\ \ P_{SS}^*=P_{PL},
\end{equation}
since the expression $H_2$ does not depend on the control variable $P_{SS}$. In accordance with the results (without the SSS) shown previously, the following analysis for the present case of SSS is carried out individually for all the four cases presented in \eqref{eq:uoptphi}. 
\paragraph{$\lambda \in [0,\infty)$} Due to the fact that $H_1$ is monotonically decreasing as $P_{SS}$ increases (for example, consider that $\frac{\partial H_1}{\partial P_{SS}}<0$) and $P_{SS}<P_{PL}$, it can be inferred that $H_1 > H_2$ for all feasible $P_{SS}$. Therefore, the associated optimal control solution shown in \eqref{eq:uoptphi}
remains as  the optimum under such circumstances, and the optimal Hamiltonian, $H^*$, is 
\begin{equation}
H^*(P_{SS}^*) \!=\! \left\{
\begin{array}{ll}
\!\!H_1^*\!=\!H_1(P_{SS_{\max}}), \!&\! \text{if} \,\,\, P_{PL}> P_{SS_{\max}}\\
\!\!H_2^*, \!& \! \text{if} \,\,\, P_{SS_{\max}} \geq P_{PL}
\end{array}\right.
\label{eq:optimalHcase1}		
\end{equation}
\paragraph{$\lambda \in (-\alpha_f Q_{\max} \eta_{dc} V_{oc},0)$} The minimum of $H_1$ by applying the  unconstrained optimal control input $P_{SS}^{*+}$ (defined in \eqref{eq:optimalpss1}) is:
\begin{multline}
H_1^* = 
q_{f0}+\alpha_f P_{PL}-\frac{\alpha_f \eta_{dc}V_{oc}^2}{4 R_b}-\frac{\lambda^2}{4 R_b\,Q_{\max}^2\alpha_f\eta_{dc}}\\-\lambda \displaystyle \frac{V_{oc}}{2 R_b\,Q_{\max}}.
\end{multline}
The minimum of the switching Hamiltonian $H$ may be identified by assuming
$
\tilde{H} = H_1^* - H_2^*,
$
which is a quadratic and concave function with respect to $P_{PL}$ ($H_1^*(P_{PL})$ is linear and $H_2^*(P_{PL})$ is convex). The maximum value of $\tilde{H}$, $\tilde{H}_{max}$, can be identified by first solving the equation ${\partial \tilde{H}}/{\partial P_{PL}} = 0$, yielding 
$
P_{PL}^* = P_{SS}^{*+}>0
$
from which it is immediate to obtain $\tilde{H}_{max} = \tilde{H}(P_{PL}^*) = q_{f0}>0$. Hence, the equation $\tilde{H} = 0$ has two real roots $P_{PL}^+(\lambda)$ and $P_{PL}^-(\lambda)$ (the superscript ``$-$'' stands for the smaller root while $``+''$ represents the greater one), which result in a power interval $\Sigma_1 = \{P_{PL}|P_{PL}^-<P_{PL}<P_{PL}^+\}$, and for $P_{PL} \in \Sigma_1$, it holds that
\begin{equation}
\tilde{H} > 0 \Rightarrow H_1^* > H_2^*\,.
\label{eq:H1greatH2}
\end{equation}
Expression \eqref{eq:H1greatH2} gives the region where $H_2^*$ is the minimum of the switching Hamiltonian. Otherwise, $H_1^*$ is the minimum. A singularity may occur when $P_{PL}$ equals $P_{PL}^+$ or $P_{PL}^-$ as both minimum candidates adopt the same value. In this case, either $H_1^*$ or $H_2^*$ can be selected, and the preference could depend on the emphasis placed on other aspects, including NOx emissions, driver comfort, engine noise, and SSS cost (response lag and additional fuel required to restart the engine). The present work focuses 
on the fuel efficiency of energy management, which is directly influenced by the engine start fuel cost (as modeled in \eqref{eq:hybridsys1}-\eqref{eq:hybridsys2}). In this context, $H_1^*$ should be chosen at the singularity to minimize the number of engine on/off switches. 

If $P_{PL}>P_{SS_{\max}}$, it is straightforward to show that the Hamiltonian is minimised at $H_1^*$, as $H^*=H_2^*$ is valid only when $P_{PL} \leq P_{SS_{\max}}$ (see \eqref{eq:optimalHcase1}). Let us now consider the case $P_{PL} \leq P_{SS_{\max}}$. When the unconstrained optimal control, $P_{SS}^{*+}$, is greater than $P_{PL}$ (that is equivalent to $P_{PL}<P_{PL}^-$), $P_{SS}^*$ will be saturated by $P_{PL}$ such that $P_{SS}^*=P_{PL}$. Hence, the $P_{PL}$ power region that yields $P_{SS}^* = P_{PL}$ (i.e., $H^*=H_2^*$, full electric mode) is $\mathcal{P}_{e,1} \triangleq \{P_{PL}|\Sigma_1 \cup P_{SS}^{*+} > P_{PL} \cap P_{PL} \leq P_{SS_{\max}}\}$.
Therefore, the optimal solution for $\lambda \in (-
\alpha_f Q_{\max} \eta_{dc} V_{oc},0)$, can be expressed as: 
\begin{equation}
H^*(P_{SS}^*) \!=\! \left\{
\begin{array}{ll}
\!\!H_2^*, & P_{PL}\in \mathcal{P}_{e,1},\\
\!\!H_1^*=H_1(\min(P_{SS}^{*+},P_{SS_{\max}})), & \text{otherwise}.
\end{array}\right.
\label{eq:optimalHcase2}		
\end{equation}

By following in this subsection the same steps conducted previously, it is straightforward to derive the $P_{PL}$ regions where $H_1^* > H_2^*$ in the remaining two scenarios shown in \eqref{eq:uoptphi}:  c) $\lambda \in \left[\displaystyle -{\alpha_f Q_{\max} V_{oc}}/{\eta_{dc}},-\alpha_f Q_{\max} \eta_{dc} V_{oc}\right]$ and d) $\lambda \in \left(-\infty, \displaystyle -{\alpha_f Q_{\max} V_{oc}}/{\eta_{dc}}\right)$. Without loss of generality, let us assume the condition \eqref{eq:H1greatH2} is valid in case c) for $P_{PL} \in \mathcal{P}_{e,2}$, and in case d) for $P_{PL} \in \mathcal{P}_{e,3}$. Thus, the optimal solution in both scenarios can be expressed as:
\begin{equation}\label{eq:uoptphi3}
H^*(P_{SS}^*) \!=\!
\left\{\begin{array}{lr}
\!\!\!H_2^*, &  P_{PL}\in \mathcal{P}_{e,2}\,,\\
\!\!\!H_1^*=H_1(0),  & \text{otherwise}\,, 
\end{array}\right.
\end{equation}
for $\lambda \in \left[\displaystyle -{\alpha_f Q_{\max} V_{oc}}/{\eta_{dc}},-\alpha_f Q_{\max} \eta_{dc} V_{oc}\right]$
and
\begin{equation}\label{eq:uoptphi4}
H^*(P_{SS}^*) \!\!=\!\!
\left\{\begin{array}{l}
\!\!\!\!H_2^*, \hspace{4.2cm} P_{PL}\in \mathcal{P}_{e,3}\,,\\
\!\!\!\!H_1^*\!=\!H_1\!\left(\max\left( P_{SS}^{*-},P_{PL}\!-\!P_{PS_{\max}},P_{SS_{\min}}\right)\right), \\ \hspace{5cm} \text{otherwise}\,, 
\end{array}\right.
\end{equation}
for $\lambda \in \left(-\infty, \displaystyle -{\alpha_f Q_{\max} V_{oc}}/{\eta_{dc}}\right)$.
\begin{figure}[htp!]
\centering
\includegraphics[width=\columnwidth]{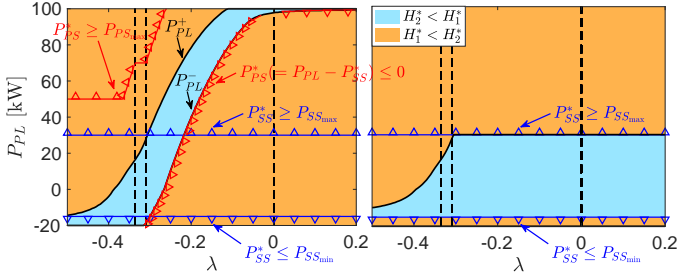}\\[-2ex]
\caption{A graphical representation of $H^*$ for the HEV parameters given in Table~\ref{tab:vehicle data}. Orange color represents regions where $H_1^*$ is the minimum and blue color represents regions where $H_2^*$ is the minimum. The black solid lines represent the non-unicity solutions where $H_1^* = H_2^*$, while the vertical black dashed lines from left to right are $\lambda = -{\alpha_f Q_{\max} V_{oc}}/{\eta_{dc}},\,\lambda =-\alpha_f Q_{\max} \eta_{dc} V_{oc},\lambda =0$, which define different costate regions. The left plot shows the solution without involving control constraints \eqref{eq:OCPconstraint5}(or \eqref{eq:OCPconstraint2}-\eqref{eq:OCPconstraint3}), although constraints are displayed as sawtooth lines. The sawtooth facing regions are invalid once the PS and SS power constraints are included.
The right plot shows the minimum Hamiltonian solution combining the control constraints displayed in the left plot.}
\label{fig:graphic_H1H2}
\end{figure}

A graphical representation of the optimal solutions is illustrated in Fig.~\ref{fig:graphic_H1H2} for different values of costate $\lambda$ (horizontal-axis) and load power $P_{PL}$ (vertical-axis). As it can be seen, the optimum $H^*$ selects the minimum among the two candidates, $H_1^*$ and $H_2^*$ (with the regions of different color in Fig.~\ref{fig:graphic_H1H2} specifying where each candidate has the lowest value), and when $H_1^*=H_2^*$, $H_1^*$ is selected under the provision of minimizing ICE start-stop events. The left plot in Fig.~\ref{fig:graphic_H1H2} denotes the solution without including the control constraints on $P_{SS}$ (and $P_{PS}$ as a consequence). In this case the two solution regions are separated by two borders (shown as thick solid lines), on which $\tilde{H} = 0$ (singularity). In particular, for $\lambda \in (-\alpha_f Q_{\max} \eta_{dc} V_{oc},0)$, the upper and lower borders respectively denote $P_{PL}\!=\!P_{PL}^+$ and $P_{PL}\!=\!P_{PL}^-$ as they have been previously defined in Section~\ref{subsec:fundamental2}-b). The left plot in Fig.~\ref{fig:graphic_H1H2} is also overlaid with lines that show the control constraints (obtained by \eqref{eq:OCPconstraint2}-\eqref{eq:OCPconstraint3} or \eqref{eq:OCPconstraint5}). By including these control constraints in the problem, the practical solution is shown in the right plot of Fig.~\ref{fig:graphic_H1H2}, which verifies the closed form solution described by \eqref{eq:optimalHcase1}, \eqref{eq:optimalHcase2}, \eqref{eq:uoptphi3} and \eqref{eq:uoptphi4}.

In conclusion, the optimal solution $P_{SS}^*$ that minimizes the piecewise Hamiltonian \eqref{eq:hamiltonian2} is formed by the four segments, as shown in \eqref{eq:optimalHcase1},	\eqref{eq:optimalHcase2}, \eqref{eq:uoptphi3} and \eqref{eq:uoptphi4}. Each segment is a piecewise function that merges a section of \eqref{eq:uoptphi}, obtained in absence of the SSS, with $P_{SS}^* =P_{PL}$ (except for $\lambda\geq 0$ where the control law is unique(see the first line of \eqref{eq:uoptphi} and \eqref{eq:optimalHcase1}). Depending on the PL branch power demand $P_{PL}$, the optimal control policy may switch between the two pieces of the solution within one segment. 
As with the previous no SSS case, the unconstrained closed form solution can be found explicitly by numerically identifying $\lambda$, which determines the switching threshold and the optimal SS operating power.
If a state constraint is violated in the unconstrained solution, the recursive algorithm described in Section~\ref{para:activeconstaint} can be utilized to iteratively find the optimal solution. 

\begin{figure}[htb!]
\centering
\includegraphics[width=.95\columnwidth]{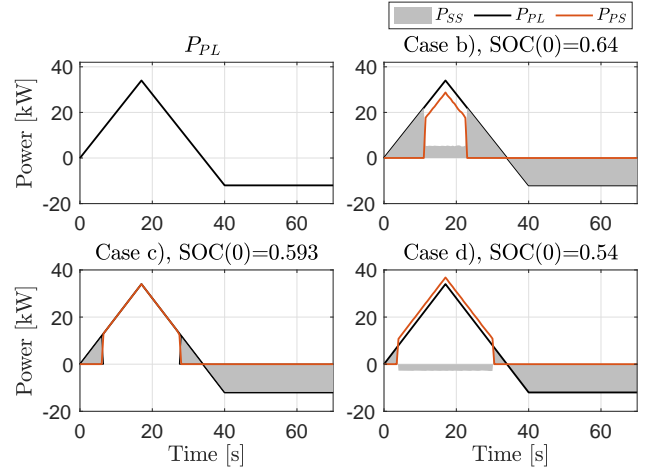}\\[-2ex]
\caption{Optimal power-split solutions found by DP for the case with the lossless SSS for an example power demand trajectory $P_{PL}$ of 70\,s, an $\soc(T)=0.65$ and three different cases of $\soc(0)$. Each $\soc(0)$ case is chosen to respectively actualize the closed-form solution cases in \eqref{eq:optimalHcase2}-\eqref{eq:uoptphi4}.
The power interval where the powertrain is operated in full electric mode for each solution is $\Sigma_1\!=\! \left[0,21.98\right]$~kW, $\Sigma_2\!=\! \left[0,12.47\right]$~kW and $\Sigma_3\!=\! \left[0,6.987\right]$~kW, respectively.}
\label{fig:example_withSSS}
\end{figure}

Similarly to the previous case without the SSS in Fig. \ref{fig:example_withoutSSS}, example DP solutions for the present case are illustrated in Fig.~\ref{fig:example_withSSS} for the same $P_{PL}$ profile, vehicle parameters and target $\soc(T)$ at $0.65$. Figure~\ref{fig:example_withSSS} shows that the three representative analytic solutions cases \eqref{eq:optimalHcase2}-\eqref{eq:uoptphi4} (\eqref{eq:optimalHcase1} is not shown since it is a sub-solution of \eqref{eq:optimalHcase2}) are precisely followed. The optimal power profiles indicate that during the propulsive phase $\Phi$ the powertrain is operated in pure electric mode at low load requirements. Once $P_{PL}$ reaches the switching threshold, the PS is activated and the powertrain is operated in ICE only or hybrid mode with the SS being charged/discharged at a constant power as with the behaviour in no SSS case.

The presence of the switching threshold therefore represents a fundamental difference of the present case solution to the solution of the case with no SSS, derived in Section \ref{subsec:fundamental1}. With no SSS, only one parameter is needed to reconstruct the solution, the constant power at which the SS power levels off once the ICE is on; for example, in Cases 2, 3, and 4 in Fig. \ref{fig:example_withoutSSS}, it levels off respectively at a positive value, zero, and a negative value; see also the parameter $P_{SS,th}$ in the context of Case 2 in \cite{boli:ecc19}. In contrast, with the SSS present, two parameters are required to reconstruct the solution, the constant power at which the SS power levels off once the ICE is on, as before, and the switching threshold.


%
%

\subsection{Hysteresis power threshold strategy (\emstwo)}
By using insights gained from the preceding solutions presented in Sections~\ref{subsec:fundamental1} and \ref{subsec:fundamental2}, the {\emstwo} is developed in this section to address the most realistic case, where the penalty fuel for the engine reactivation is involved, and for which analytic solutions are not feasible. 

\begin{figure}[htb!]
\centering
\includegraphics[width=\columnwidth]{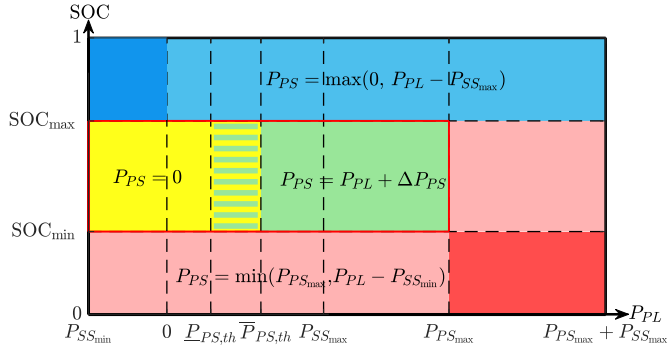}\\[-2ex]
\caption{The operation scheme of the {\emstwo} with different operating stages classified based on the given SOC and $P_{PL}$. The primary operation mode is active in the region where the coordinates (on the $P_{PL}$-SOC map) satisfy $\soc_{\min} < \soc < \soc_{\max}$ and $P_{SS_{\min}}\leq P_{PL} \leq P_{PS_{\max}}$, which give rise to a rectangular area designated by a red solid frame. $\Delta P_{PS}$ is a tuneable parameter. The hysteresis zone is shown by the mixture of yellow and green colors. Red and blue zones represent emergency handling operations\protect\footnotemark.}
\label{fig:ControlScheme}
\end{figure}

When engine start fuel cost, $m_{p}$ is enabled, $m_{p}{=}K q_{f0}$ is added to the base fuel consumption as long as the transition $\mathcal{S}_{0 \rightarrow 1} \triangleq \{s|s\!=\!0,\,s^+\!=\!1\}$ is detected. Application of the optimal EM solution derived 
for a lossless SSS to this case may result in fast engine on/off switching dynamics when $P_{PL}$ fluctuates around switching power thresholds, thus leading to a significant increase of fuel usage.
The {\emstwo} attempts to address this issue and to approximate the global optimal solution with consideration of $m_p$ by combining the control policies extracted from the analytic solutions obtained previously with a newly designed switching logic for ICE on/off control.
The overall control scheme is graphically shown in Fig.~\ref{fig:ControlScheme}.
\footnotetext{Operation in the lower right and upper left corners, shaded as red and blue regions, by the defined rules is only possible in transient conditions (short time) to avoid draining or overflowing the SOC. The latter can be naturally avoided by assigning more mechanical brakes so that, for example, $P_{PL}=0$ (see \eqref{eq:PPL2}).}

As it can be noticed, the principles of {\emstwo} are defined based on a 2-dimensional map of the SOC and $P_{PL}$, which is partitioned into several zones by the SOC limits and power load thresholds. 
Similarly to the analytic solutions derived based on the lossless SSS, $P_{PL}$ is followed by $P_{SS}$ only at low power loads (including negative $P_{PL}$) and the PS is activated, in hybrid or PS-only mode, at higher load requirements. More specifically, when the ICE is activated, it is operated at $P_{PS} = P_{PL} + \Delta P_{PS}$ with $\Delta P_{PS}$ a tuneable constant parameter, thus the SS is always operated at a constant power $-\Delta P_{PS}$ when the ICE is active. This operation is inspired by the analytic solution \eqref{eq:uoptphi}, \eqref{eq:optimalHcase2}-\eqref{eq:uoptphi4} (in which $\Delta P_{PS}$ is in fact a constant that depends on $\Delta \soc$ and is found analytically), and it introduces an additional degree of freedom (the tuneable parameter $\Delta P_{PS}$) as compared to the conventional load following (exclusive operation) strategy ($\Delta P_{PS}=0$) used in \cite{wassif:2016}. Depending on the sign and value of $\Delta P_{PS}$, when the ICE is activated the SS may be discharged to cover the unfulfilled power demand ($\Delta P_{PS}<0$), charged by the PS to absorb the excess PS power ($\Delta P_{PS}>0$), or idling ($\Delta P_{PS}=0$ and the mode falls into the PS-only mode). 

\begin{figure}[htb]
  \centering
  \input{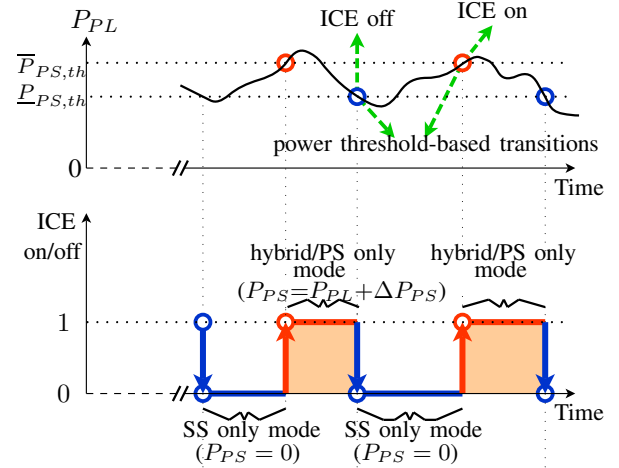}
  \vspace{-1cm}
  \caption{Hysteresis switching scheme for ICE on/off control.}
 \label{fig:switching scheme}
\end{figure}

In order to reduce the incidence of ICE on/off transitions, a hysteresis switching scheme for ICE on/off control is also developed, as illustrated in Fig.~\ref{fig:switching scheme}. 
The hysteresis dynamics are assigned in relation to the engine on/off state $s$, giving, 
\begin{equation}\label{eq:hpfs}
P_{SS} = (1-s) P_{PL}-s\Delta P_{PS},\,\,P_{PS} = s (P_{PL} + \Delta P_{PS}),
\end{equation}
and
\begin{equation*}
s = \left\{
\begin{array}{lll}
0, &   \text{if } P_{PL} \leq \underline{P}_{PS,th}\\
1, & \text{if } P_{PL} > \overline{P}_{PS,th}\\
s(t^-), &\text{otherwise},
\end{array}\right.
\end{equation*}
where $t^-$ represents the time instant before $t$, and $\underline{P}_{PS,th}$ and $\overline{P}_{PS,th}$ are two separate tuneable power thresholds. The hysteresis dynamics dictate the operation in the hysteresis zone in Fig. \ref{fig:ControlScheme}, while it is clear that outside the hysteresis zone the $P_{SS}$ and $P_{PS}$ expressions in \eqref{eq:hpfs} fall back respectively to those in the yellow and green zones in Fig. \ref{fig:ControlScheme}.

Operation outside the primary region triggers the emergency rules, which aim to prevent the SOC constraints violation (which may happen in practice due to control discretization) and also define the power split for extremely large power demand ($P_{PL}>P_{PS,\max}$). More specifically, when SOC reaches or goes beyond its limits ($\soc \geq\soc_{\max} \vee \soc \leq \soc_{\min}$), the SS power is set to maximum ($\min(P_{SS_{\max}},P_{PL})$) or minimum ($\max(P_{PL}\!-\!P_{PS_{\max}},\,P_{SS_{\min}})$) operating power, respectively, to force the SOC immediately back into the main operational zone, as inspired by the PMP analysis in Section~\ref{para:activeconstaint} and Fig.~\ref{fig:discontinued_example}. In particular, with the operational policy for $\soc \leq \soc_{\min}$, the PS can be triggered on (enabling the hybrid mode) even during braking.

The design of the {\emstwo} amounts to finding optimally tuned values for  $\underline{P}_{PS,th}$, $\overline{P}_{PS,th}$ and $\Delta P_{PS}$ that minimize the fuel consumption $m_f$ while maintaining the CS condition:
\begin{equation}\label{eq:cscondition}
\Delta \soc = 0.
\end{equation}
Condition \eqref{eq:cscondition} is sought because it is naturally optimal as will be shown next in Section~\ref{subsec:EFC}, which is also a major contribution of this paper.
The global tuning nature of the control requires access to the whole driving cycle in advance (as is the case with DP and ECMS) so that the tunable parameters can be tuned separately for each driving cycle. This allows for the rules of the {\emstwo} to be tailored for all driving cycles rather than having compromised control policies that may excel on some driving cycles but behave less well on others. Thus, {\emstwo} can be used to obtain benchmark solutions, and also it can be implemented in practice when the driving profile is known or can be estimated.   

%

\subsection{Fuel Economy Evaluation}\label{subsec:EFC}
The equivalent fuel consumption (EFC) is a measure of the fuel economy that has been widely used in the literature for evaluating overall fuel economy. It allows the comparison of the overall fuel economy by considering the actual fuel consumption as well as the shortage/surplus of final SOC. In this subsection we prove that the optimal EFC of a driving mission is 
achieved for CS operation, which provides additional justification beyond practical reasons about why CS operation should be sought. The definition of the EFC is \cite{Sciarretta+Back+Guzzella/IEEE:2004}: 
\begin{equation}
m_{efc}{=} 
\left\{\begin{array}{lll} 
\!\!m_f \!+\! S_{d,efc}\,\Delta \soc \displaystyle \frac{Q_{\max} V_{oc}}{q_{HV}}, \,\,\,\Delta \soc \!\geq\! 0 ,\\
\!\!m_f \!+\! S_{c,efc}\,\Delta \soc \displaystyle \frac{Q_{\max} V_{oc}}{q_{HV}}, \,\,\,\Delta \soc \!<\! 0,
\end{array}\right.
\label{eq:qefc}
\end{equation}
where the two equivalence factors $S_{d,efc}$ and $S_{c,efc}$ (for battery discharging and charging respectively) represent the correlation of the electrical energy and the fuel chemical energy required when following a driving cycle. Hence, to proceed with the assessment of an EMS, $S_{d,efc}$ and $S_{c,efc}$ have to be identified a priori for each driving cycle and for each vehicle model. In brief, the identification method proposed in \cite{Sciarretta+Back+Guzzella/IEEE:2004} requires a sweep of the power sharing factor $u_{efc}\triangleq P_{PS}/P_{PL},\,\forall t\in \Phi $ within the range $[1-\Delta{u}_{efc},\,1+\Delta{u}_{efc}]$, with $\Delta{u}_{efc}$ selected such that either the upper or the lower bound for the SOC is not violated during the operation.
The overall electrical and fuel energy consumptions for a specific value of $u_{efc}$ while undergoing a given drive cycle are respectively computed by
\begin{align*}
  &   E_e\triangleq \int_0^T i_b V_{oc}dt \,\,\,\,\,\,\,\, \text{and} & E_f\triangleq \int_{0}^{T}q_{HV}\dot{m}_f dt,
\end{align*}
and are plotted against each other for different values of $u_{efc}$. Such a plot is separated into two segments intersecting at $u_{efc}=1$, at which the propulsion power is purely provided by the ICE. The slopes of straight lines that fit the ($E_e$, $E_f$) data of these two segments are identified as the negative values of $S_{d,efc}$ and $S_{c,efc}$ respectively. It will now be shown that the EFC definition inherently drives the optimal EFC solutions (of an EMS) to be strictly CS as in \eqref{eq:cscondition}. 

As the hybrid mode is enabled only during an emergency when $P_{PL}<0$, it is reasonable to assume $P_{SS}=P_{PL},\forall t\in \Psi$. As such, the fuel consumption at the end of the driving mission, in light of  \eqref{eq:simplemodel1}, \eqref{eq:simplemodel2} and \eqref{eq:powerbalance}, is expressed as:
\begin{equation}\label{eq:overallfuel2}
m_f \!=\! q_{f0} \int_{0}^{T}  \!\!s \,dt + \alpha_f\int_{\Phi} P_{PL}dt- \alpha_f\int_{\Phi}  P_{SS} \,dt + N_r m_{p},    
\end{equation}
where $N_r$ is the number of engine restarts during the mission and $\alpha_f\int_{\Phi} \! P_{PL}dt$ is fixed for a given driving cycle and independent of the EM control. The fuel energy for a given driving cycle is $E_f=q_{HV}m_f$.
In terms of the electrical energy $E_e$, when $u_{efc} \geq 1$ it means that $E_e$ is never used for propulsion, that is $\Phi_d = \emptyset,\,\Phi = \Phi_c$, and when $u_{efc} < 1$ it is clear that $\Phi_c = \emptyset,\,\Phi = \Phi_d$. Therefore, $E_e$ can be rewritten as,
\[
E_e = \left\{
\begin{array}{lll}
\displaystyle \int_{\Phi_d} \frac{P_{SS}V_{oc}}{\eta_{dc}V_b}dt + E_{e,\Psi},\,\,\, \text{if}\,\, u_{efc} < 1,\\
\displaystyle \int_{\Phi_c} \frac{P_{SS}\eta_{dc}V_{oc}}{V_{b}}dt+ E_{e,\Psi},\, \,\, \text{if}\,\, u_{efc} \geq 1,
\end{array} \right.
\]
where $E_{e,\Psi} {=} \Delta_{\Psi} \soc\, Q_{\max}\, V_{oc}$ only depends on $P_{PL}$.
Consider two arbitrary values of $u_{efc} \geq 1 $ within the admissible set $[1,\,1+\Delta u_{efc}]$. It is obvious that $N_r$ and $s$ are invariant between the two scenarios. Then, the slope corresponding to $S_{d,efc}$ is evaluated by,
\begin{multline*}
S_{d,efc} = - \frac{\Delta E_f}{\Delta E_e} = \frac{q_{HV} \alpha_f \int_{\Phi_c} (P_{SS,1} - P_{SS,2})dt}{\eta_{dc}V_{oc}\int_{\Phi_c} \left(\frac{P_{SS,1}}{V_{b,1}} - \frac{P_{SS,2}}{V_{b,2}}\right) dt}\\
= \frac{q_{HV} \alpha_f \int_{\Phi_c} ((i_{b,1}-i_{b,2})(V_{oc} - R_b(i_{b,1}+i_{b,2}))) dt}{V_{oc}\eta_{dc}^2\int_{\Phi_c} \left({i_{b,1}} - {i_{b,2}}\right) dt},
\end{multline*}
where the subscripts 1 and 2 indicate the two scenarios driven by the two distinct $u_{efc}$ values. 
Since $V_{oc} - R_b(i_{b,1}+i_{b,2})\geq V_{oc}, \forall t\in \Phi_c$, it is obtained that: 
\begin{equation}\label{eq:boundsd}
S_{d,efc} > \frac{1}{\eta_{dc}^2}{q_{HV} \alpha_f}.
\end{equation}
Similarly, $S_{c,efc}$ is evaluated as follows, with respect to two arbitrary values of $u_{efc} < 1 $ within the admissible set $[1-\Delta u_{efc},\,1)$:
\begin{equation*}
S_{c,efc} 
\!=\! \frac{q_{HV} \alpha_f \eta_{dc}^2 \int_{\Phi_d} ((i_{b,3}\!-\!i_{b,4})(V_{oc} \!-\! R_b(i_{b,3}+i_{b,4})))dt}{V_{oc}\int_{\Phi_d} \left({i_{b,3}} - {i_{b,4}}\right) dt},
\end{equation*}
which implies
\begin{equation}\label{eq:boundsc}
S_{c,efc} < \eta_{dc}^2 q_{HV} \alpha_f.
\end{equation}
Turning to the steps required for the calculation of $m_{efc}$ and by applying \eqref{eq:eta_dc} to \eqref{eq:overallfuel2}, it holds that, 
\begin{multline}\label{eq:40}
m_f = q_{f0} \int_{0}^{T}  \!\!s \,dt \!+\! \alpha_f\int_{\Phi}  P_{PL}dt- \alpha_f\int_{\Phi} \eta_{dc} ^{\sign(P_b)} P_{b} \,dt \\ + N_r m_{p}   , 
\end{multline}
where the term $\alpha_f\int_{\Phi}\eta_{dc} ^{\sign(P_b)}  P_b \,dt$ can be expanded by using the definitions of $\Phi_d$ and $\Phi_c$, as follows:
\begin{equation}\label{eq:expansion}
 \alpha_f\int_{\Phi}\eta_{dc} ^{\sign(P_b)}  P_b \,dt = \alpha_f \left(\int_{\Phi_d}\eta_{dc}  P_b dt + \int_{\Phi_c}\frac{1}{\eta_{dc}} P_b dt\right).
\end{equation}
In relation to the charging/discharging intervals, let us define, 
\begin{equation}\label{eq:vbdvbc}
\begin{array}{ll}
 V_{b,d} \triangleq V_{oc} - R_b\,i_{b} \leq V_{oc}, & \forall t\in  \Phi_{d} \,,\\
 V_{b,c} \triangleq V_{oc} - R_b\,i_{b} > V_{oc}, & \forall t\in  \Phi_{c}\,.
\end{array}
\end{equation}
Then, \eqref{eq:expansion} can be expressed as,
\begin{equation}\label{eq:outcome}
\begin{aligned}
& \alpha_f \left(\int_{\Phi_d}\eta_{dc}  P_b dt + \int_{\Phi_c}\frac{1}{\eta_{dc}} P_b dt\right) \\
&=\! -Q_{\max}\alpha_f \!\left(\!\eta_{dc} \!\! \int_{\Phi_d} \!\!\!\! V_{b,d} \frac{d\text{\soc}}{dt}dt\!+\! \frac{1}{\eta_{dc}} \! \int_{\Phi_c} \!\!\!\! V_{b,c} \frac{d\text{\soc}}{dt}dt\! \right)\!,
 \end{aligned}
\end{equation}
where $i_b = -Q_{\max}\frac{d \soc }{dt}$ is applied. Owing to the mean value theorem, there exist two time instants $t_{d} \in \Phi_d$ and $t_{c} \in \Phi_c$, such that: 
\begin{align}
\int_{\Phi_d} \! V_{b,d} \frac{d\text{\soc}}{dt}dt \!=\! V_{b,d}(t_{d})\!\int_{\Phi_d}\! \frac{d\text{\soc}}{dt}dt \!=\! -V_{b,d}(t_{d}) \Delta_{\Phi_d}\soc \nonumber\\
\int_{\Phi_c}\! V_{b,c} \frac{d\text{\soc}}{dt}dt \!=\! V_{b,c}(t_{c})\!\int_{\Phi_c}\! \frac{d\text{\soc}}{dt}dt\!=\!-V_{b,c}(t_{c}) \Delta_{\Phi_c}\soc. \label{eq:outcome2}
\end{align}
By substituting \eqref{eq:40} in \eqref{eq:qefc} and applying \eqref{eq:outcome} and \eqref{eq:outcome2}, it is immediate to obtain the explicit expression of $m_{efc}$, as follows: 
\begin{equation}
m_{efc}\!\!=\!\! 
\left\{\begin{array}{lll}
  \!\!\!\!\displaystyle m_{f0}\! \!-\!\! \left(\!\eta_{dc}  V_{b,d}(t_{d}) \Delta_{\Phi_d}\soc \!+\! \frac{1}{\eta_{dc}} V_{b,c}(t_{c}) \Delta_{\Phi_c}\soc \!\right)\\\times Q_{\max}\alpha_f   + S_{d,efc}\Delta \text{\soc} \displaystyle\frac{Q_{\max}V_{oc}}{q_{HV}},\,\, \Delta \text{\soc} \!\geq\! 0 \,,\\
\!\!\!\!\displaystyle m_{f0}  \!\!-\!\!  \left(\!\eta_{dc}  V_{b,d}(t_{d}) \Delta_{\Phi_d}\soc \!+\! \frac{1}{\eta_{dc}} V_{b,c}(t_{c}) \Delta_{\Phi_c}\soc \!\right)\\\times Q_{\max}\alpha_f  + S_{c,efc}\Delta \text{\soc} \displaystyle \frac{Q_{\max}V_{oc}}{q_{HV}},\,\, \Delta \text{\soc} \!<\! 0\,,
\end{array}\right.
\label{eq:mefc_analytical}
\end{equation}
where, 
$
m_{f0} = q_{f0} \int_{0}^{T} \! s \,dt +\! N_r m_{p}+\alpha_f \int_\Phi P_{PL} \,dt,
$
and the piecewise function \eqref{eq:mefc_analytical} is continuous at $\Delta \text{\soc}=0$.

By using \eqref{eq:DeltaSOC}, \eqref{eq:mefc_analytical} can be rearranged into \eqref{eq:mefc_analytical3}. 
\begin{figure*}
\begin{equation}
m_{efc} \!=\! 
\left\{\begin{array}{lll}
  \displaystyle \frac{Q_{\max}V_{oc}}{q_{HV}}\left(S_{d,efc}- \eta_{dc}\alpha_fq_{HV}\frac{V_{b,d}(t_d)}{V_{oc}} \right)\Delta_{\Phi_d} \soc  +
  \frac{Q_{\max}V_{oc}}{q_{HV}}\left(S_{d,efc}- \frac{1}{\eta_{dc}}\alpha_fq_{HV}\frac{V_{b,c}(t_c)}{V_{oc}} \right)\Delta_{\Phi_c} \soc \\ \hspace{2.8cm}\displaystyle +q_{f0} \int_{0}^{T} \! s \,dt +\! N_r m_{p}+\alpha_f \int_\Phi P_{PL} \,dt + S_{d,efc}\Delta_{\Psi} \text{\soc} \displaystyle \frac{Q_{\max}V_{oc}}{q_{HV}},\,\, \Delta \soc \geq 0 \,,\\
 \displaystyle \frac{Q_{\max}V_{oc}}{q_{HV}}\left(S_{c,efc}- \eta_{dc}\alpha_fq_{HV}\frac{V_{b,d}(t_d)}{V_{oc}} \right)\Delta_{\Phi_d} \soc  +
  \frac{Q_{\max}V_{oc}}{q_{HV}}\left(S_{c,efc}- \frac{1}{\eta_{dc}}\alpha_fq_{HV}\frac{V_{b,c}(t_c)}{V_{oc}} \right)\Delta_{\Phi_c} \soc \\ \hspace{2.8cm}\displaystyle +q_{f0} \int_{0}^{T} \! s \,dt +\! N_r m_{p}+\alpha_f \int_\Phi P_{PL} \,dt+ S_{c,efc}\Delta_{\Psi} \text{\soc} \displaystyle \frac{Q_{\max}V_{oc}}{q_{HV}},\,\, \Delta \soc < 0 \,.
\end{array}\right.
\label{eq:mefc_analytical3}
\end{equation}
\hrule
\end{figure*}
As it can be seen, \eqref{eq:mefc_analytical3} is a piecewise bilinear function of $V_{b,d}(t_d)$, $V_{b,c}(t_c)$, $\Delta_{\Phi_c} \soc$, $\Delta_{\Phi_d} \soc$, $s$ and $N_r$, which are all influenced by the EM strategy through $P_{SS}$. On the other hand, the last two terms of each part in \eqref{eq:mefc_analytical3} are independent of the EM and only depend on $P_{PL}$, and therefore they are constants for a given $P_{PL}$ profile. 
As a consequence, $\Delta_{\Phi} \soc$ is determined once a pair of boundary conditions of SOC $(\soc(0),\soc(T))$ is given (that determine $\Delta \soc$). The $P_{SS}$ profile that meets the the desired $\Delta_{\Phi} \soc$ is not unique, and it is possible to find some $P_{SS}$ profile that sets $V_{b,d}, V_{b,c}$ independently of each other to some desirable profiles, and as a result, $\Delta_{\Phi_d} \soc$, $\Delta_{\Phi_c} \soc$, $s$ and $N_r$ (that are respectively determined by $V_{b,d},  V_{b,c}$) can be independently assigned to desired values. 
By considering the inequality conditions \eqref{eq:boundsd} and \eqref{eq:boundsc}, as well as \eqref{eq:vbdvbc} and that $\eta_{dc}\leq 1$, it can be inferred that:
\begin{align}
S_{d,efc}- \eta_{dc}\alpha_fq_{HV}\frac{V_{b,d}(t_d)}{V_{oc}}>0, \label{eq:sdefcineq} \\
S_{c,efc}- \frac{1}{\eta_{dc}}\alpha_fq_{HV}\frac{V_{b,c}(t_c)}{V_{oc}} < 0\,.\label{eq:scefcineq}
\end{align}
By referring to the $\Delta \soc \geq 0$ case in \eqref{eq:mefc_analytical3} it can be easily inferred that $m_{efc}$ is minimized when $\Delta_{\Phi_d} \text{\soc}$ is minimized, since \eqref{eq:sdefcineq} is true. By further taking into account \eqref{eq:DeltaSOC}, it can be concluded that $\Delta \soc = 0$ is necessary when $m_{efc}$ is minimum.
Similarly, when $\Delta \soc  <0$ in \eqref{eq:mefc_analytical3}, $m_{efc}$ is minimized when $\Delta_{\Phi_c} \text{\soc}$ is maximized, since \eqref{eq:scefcineq} is true. Therefore, by referring to \eqref{eq:DeltaSOC} the necessary condition to minimize $m_{efc}$ in this case is also $\Delta \soc = 0$ ( owing to the continuity of \eqref{eq:mefc_analytical3} at $\Delta \soc = 0$).
Hence, the strictly CS condition \eqref{eq:cscondition} is a \emph{necessary} condition overall for EFC minimization.


\section{Numerical Results}
\label{sec:simulation}

\begin{figure}[htb!]
\centering
\includegraphics[width=\columnwidth]{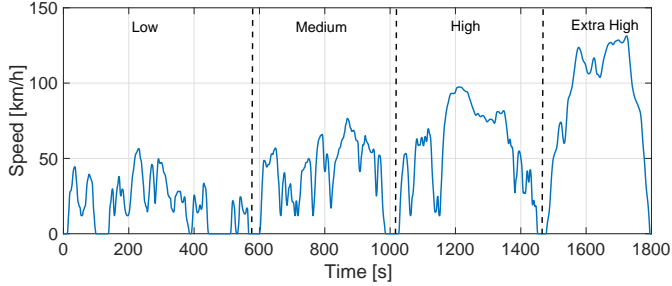}\\[-2ex]
\caption{Speed profile of WLTP with four individual stages classified by their average speed.}
\label{fig:WLTP}
\end{figure}

\begin{figure}[H]
\centering
\hspace{6mm}\includegraphics[width=0.88\columnwidth]{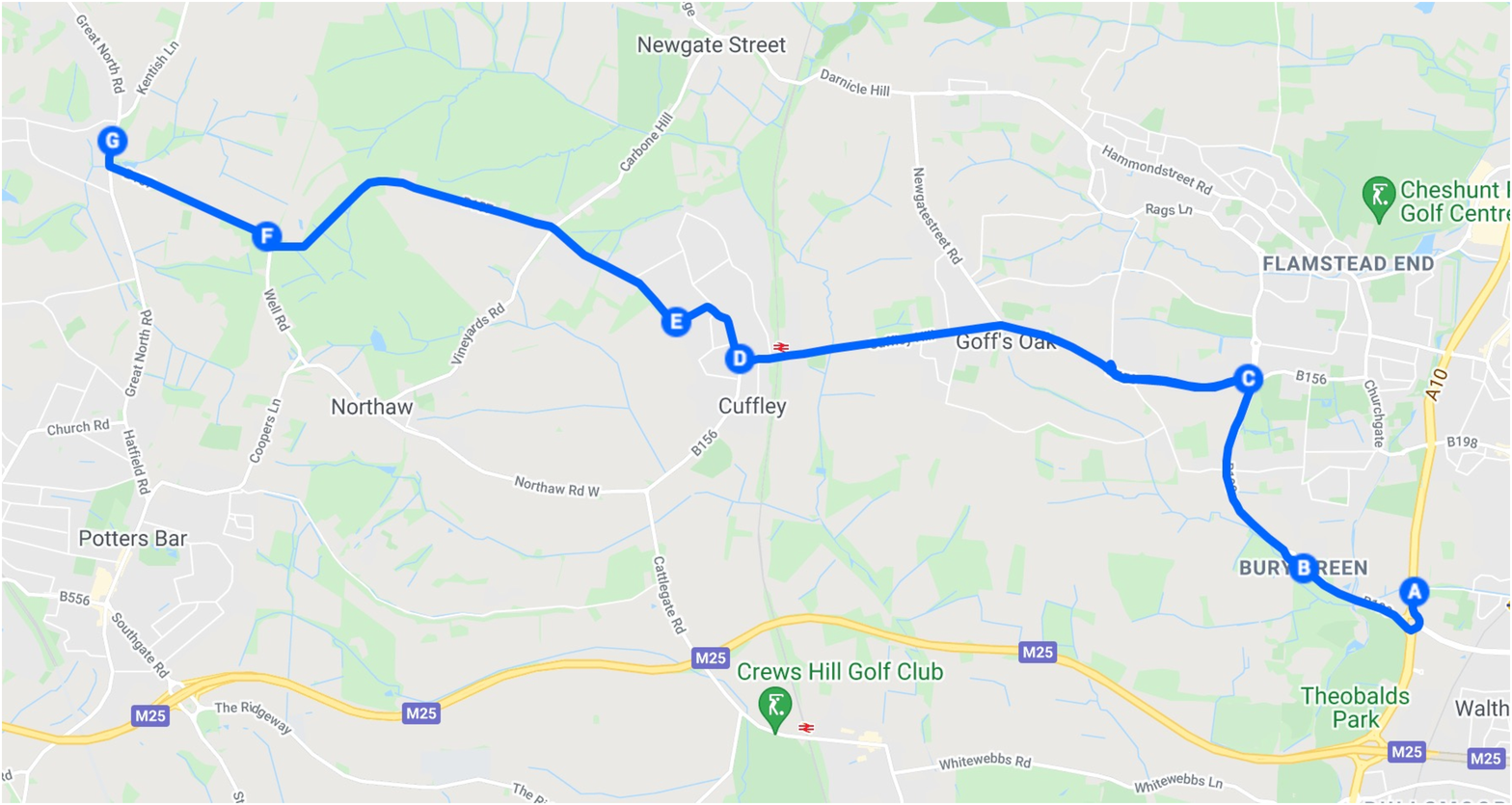}\\
\includegraphics[width=\columnwidth]{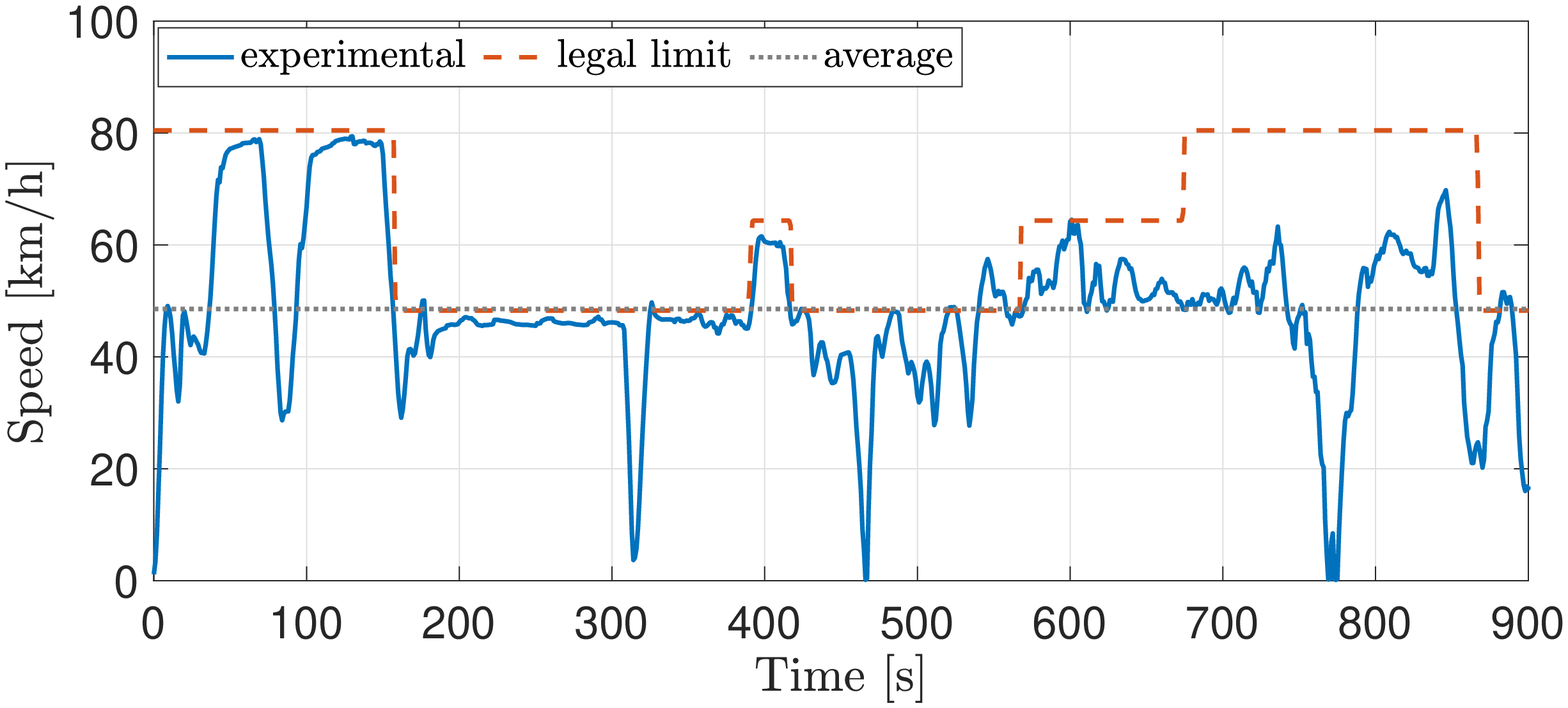}\\[-2ex]
\caption{Top: 12.1km rural route selected as a real-world vehicle driving mission https://t.ly/n2wL. Bottom: Experimental driving speed profile for the mission above.}
\label{fig:ruralLondon}
\end{figure}

The EM control strategies considered and developed in this work are tested by simulations in which the vehicle follows predefined driving cycles. The WLTP (worldwide harmonized light vehicles test procedure) corresponds to the latest test procedure adopted by industry and
it is therefore utilized in the present work. As shown in
Fig.~\ref{fig:WLTP}, the WLTP profile is a single driving cycle with four
stages, defined by their average speed: low (WL-L), medium (WL-M),
high (WL-H) and extra high (WL-E). Each of the stages can
be considered on their own as independent driving cycles.
In addition to the WLTP, an experimental speed profile (shown in Fig.~\ref{fig:ruralLondon}) is also adopted for performance and robustness assessment purposes. This time history data, that is recorded by a newly built data acquisition device \cite{yan:2017}, exhibits realistic driving behavior on a rural road. As compared to standard test cycles, this experimental speed pattern contains particular features that better reflect real-world driving, such as the influences of legal speed limits and road grades, and the driving style of the human driver who is inclined to apply higher values of acceleration and deceleration than in the WLTP.

The proposed HPTS is individually applied to the four segments of the WLTP speed profile, and the solutions are benchmarked against DP \cite{Elbert:TCST2013} and XOS \cite{wassif:2016} in the context of both linear FCM \eqref{eq:Q_f} and experimental (quasilinear) FCM (dotted line in the right plot of Fig. \ref{fig:fuelmassrate}) for robustness verification purposes. 
It is noteworthy that ECMS, which is one of the popular EM control strategies in the literature, breaks down for the problem addressed in this paper, and therefore not used for comparison. The reason is briefly explained as follows. ECMS finds the optimal power split to minimize an equivalent fuel consumption, defined as
\begin{equation}
\dot{m}_{eq}=\left\{ \begin{array}{ll}
\displaystyle \dot{m}_{f}(P_{PS}) + S_{d} \frac{P_{SS}}{q_{HV}} & P_{SS}\ge 0
 \\[1ex] 
\displaystyle \dot{m}_{f}(P_{PS}) + S_{c} \frac{P_{SS}}{q_{HV}} & P_{SS}<0  
 \end{array} ,
\right.
\label{eq:efcrate}
\end{equation}
where the two constants $S_d$ and $S_c$ are equivalence factors that translate the energy discharged/charged by the SS into a corresponding amount of fuel consumed/stored. Due to the linear/quasilinear FCM for a series HEV, \eqref{eq:efcrate} becomes a linear/quasilinear combination of $P_{PS}$ and $P_{SS}$ with individual gradients depending on $\alpha_f$, $q_{HV}$, $S_d$ and $S_c$. Hence, the ECMS simply operates $P_{PS}$ always at its maximum or always at its minimum throughout a mission, irrespective of the power demand, and with the choice of (always) maximum or minimum $P_{PS}$ depending on the sign of the gradients, unless SOC limits are reached.
For a fair comparison and also to satisfy the necessary condition of fuel consumption optimality, the same SOC CS boundary condition $\soc(0)\!=\!\soc(T)\!=\!0.65$ is imposed for all methods. The penalty fuel coefficient for engine restarts is set to $K=0.8$ in the first instance, while an investigation of its influence on the comparative results is also carried out. 

\begin{figure}[htb!]
\centering
\includegraphics[width=\columnwidth]{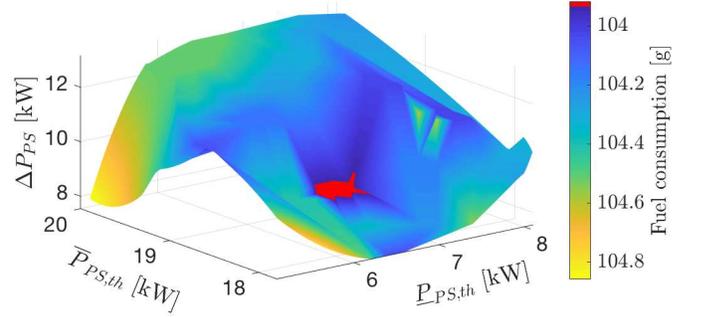}\\[-2ex]
\caption{HPTS optimal solutions obtained by tuning $\overline{P}_{PS,th}$, $\underline{P}_{PS,th}$ and $\Delta P_{PS}$, when the linear FCM model \eqref{eq:Q_f} is employed and while satisfying charge sustaining operation.}
\label{fig:HPFS_search_linear}
\end{figure}

Given a driving cycle, the proposed {\emstwo} is applied by tuning the design parameters $\underline{P}_{PS,th}$, $\overline{P}_{PS,th}$ and $\Delta P_{PS}$, while also aiming to satisfy the CS condition as mentioned. Thus, the tuning process involves finding the combination of the three parameters, among all combinations that lead to CS operation, that minimizes the fuel consumption; the EFC is now equal to the fuel consumption due to the CS condition. 
Figure~\ref{fig:HPFS_search_linear} presents an example of the tuning graph for the WL-M cycle with the linear FCM model \eqref{eq:Q_f}. 
The surface in Fig.~\ref{fig:HPFS_search_linear} denotes the control solutions satisfying $\Delta \soc\!=\!0$, and the optimal solution in terms of fuel consumption is identified approximately at $\overline{P}_{PS,th}\!=\!18.5$\,kW, $\underline{P}_{PS,th}\!=\!6.5$\,kW and $\Delta P_{PS}\!=\!9.5$\,kW. 

\begin{figure}[htb!]
\centering
\includegraphics[width=\columnwidth]{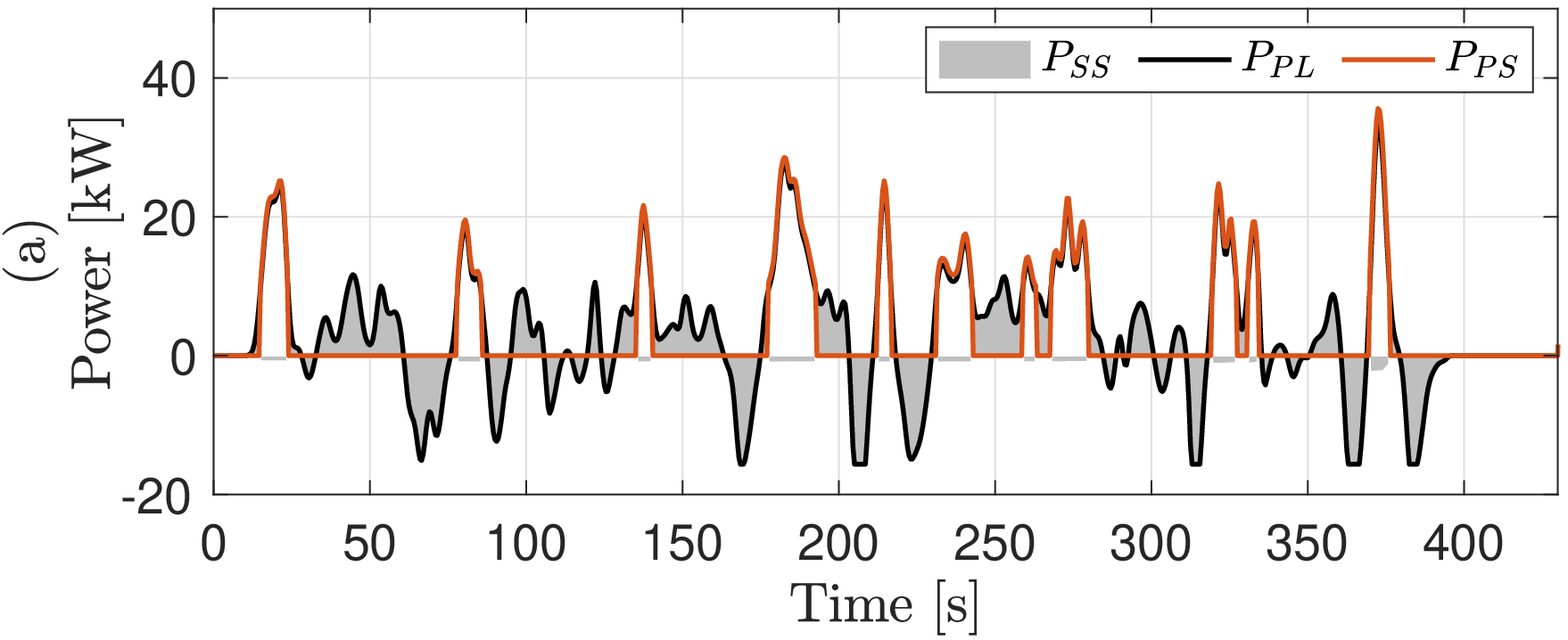}\\[-4.3ex]
\includegraphics[width=\columnwidth]{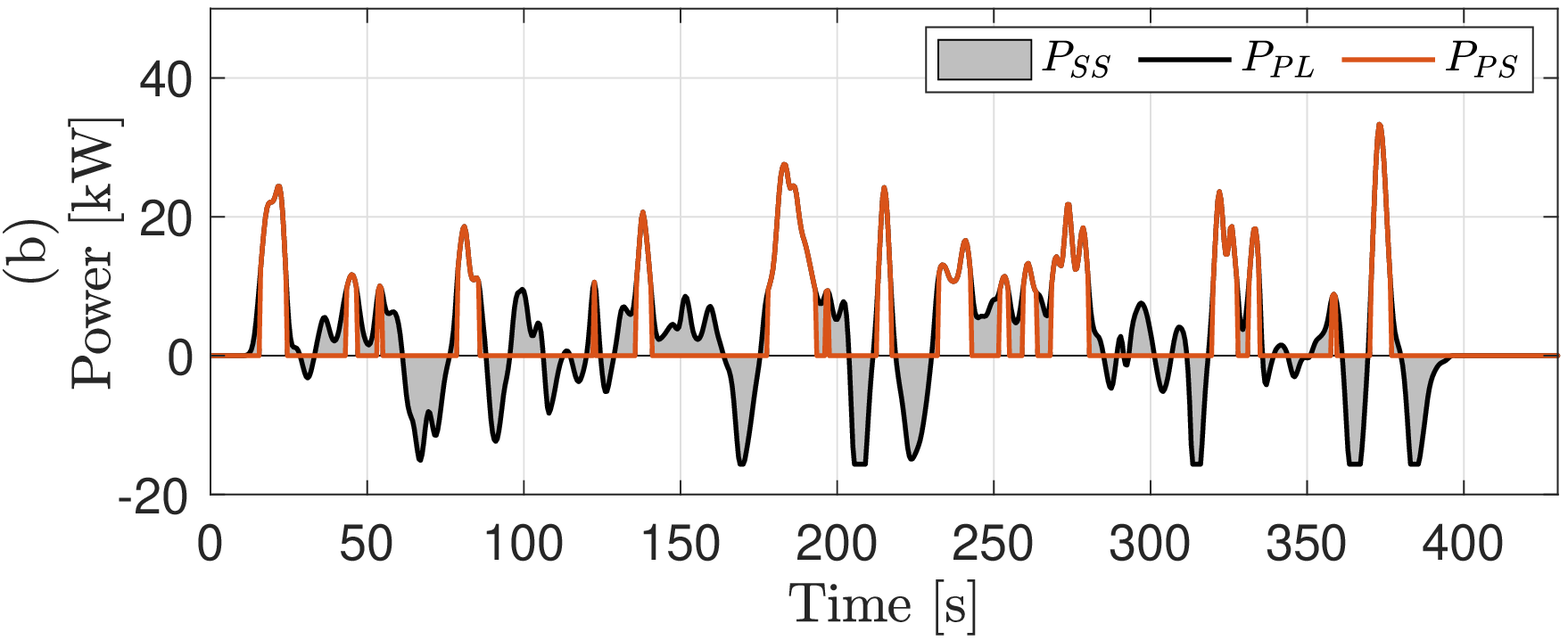}\\[-4.3ex]
\includegraphics[width=\columnwidth]{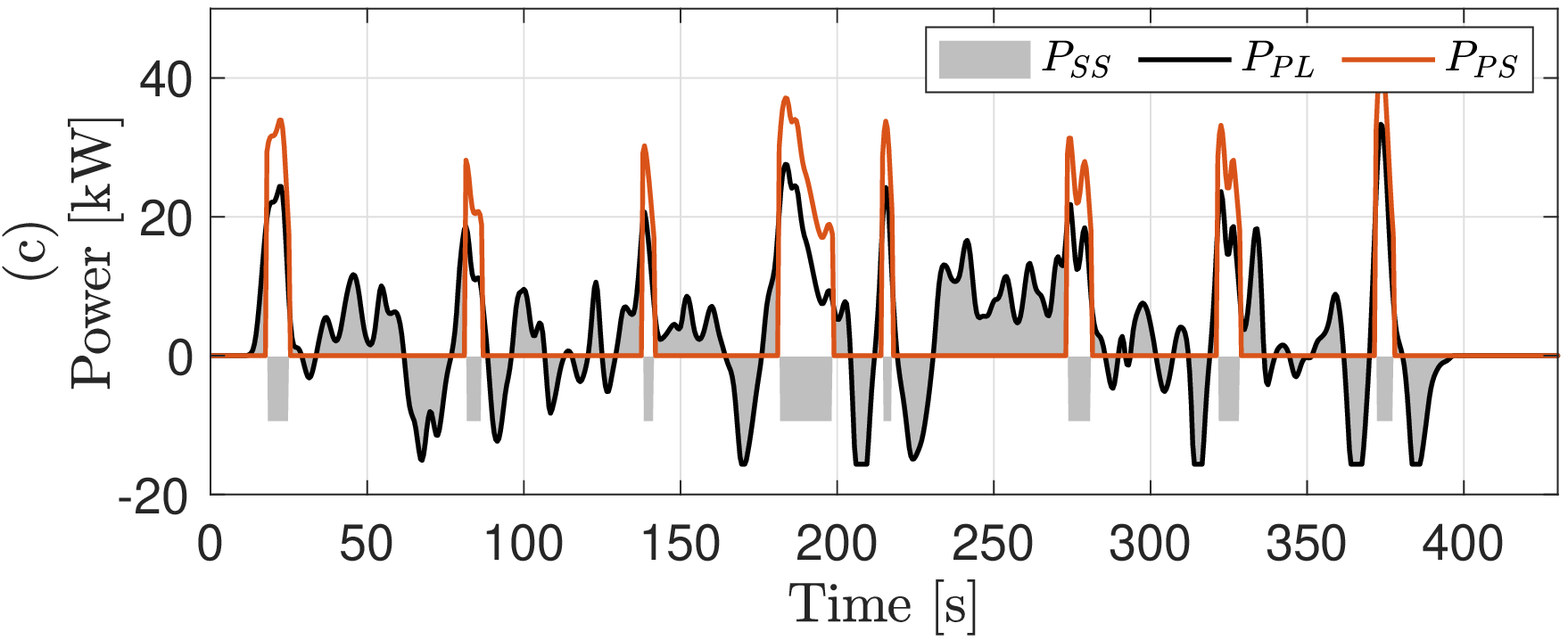}\\[-4.3ex]
\includegraphics[width=\columnwidth]{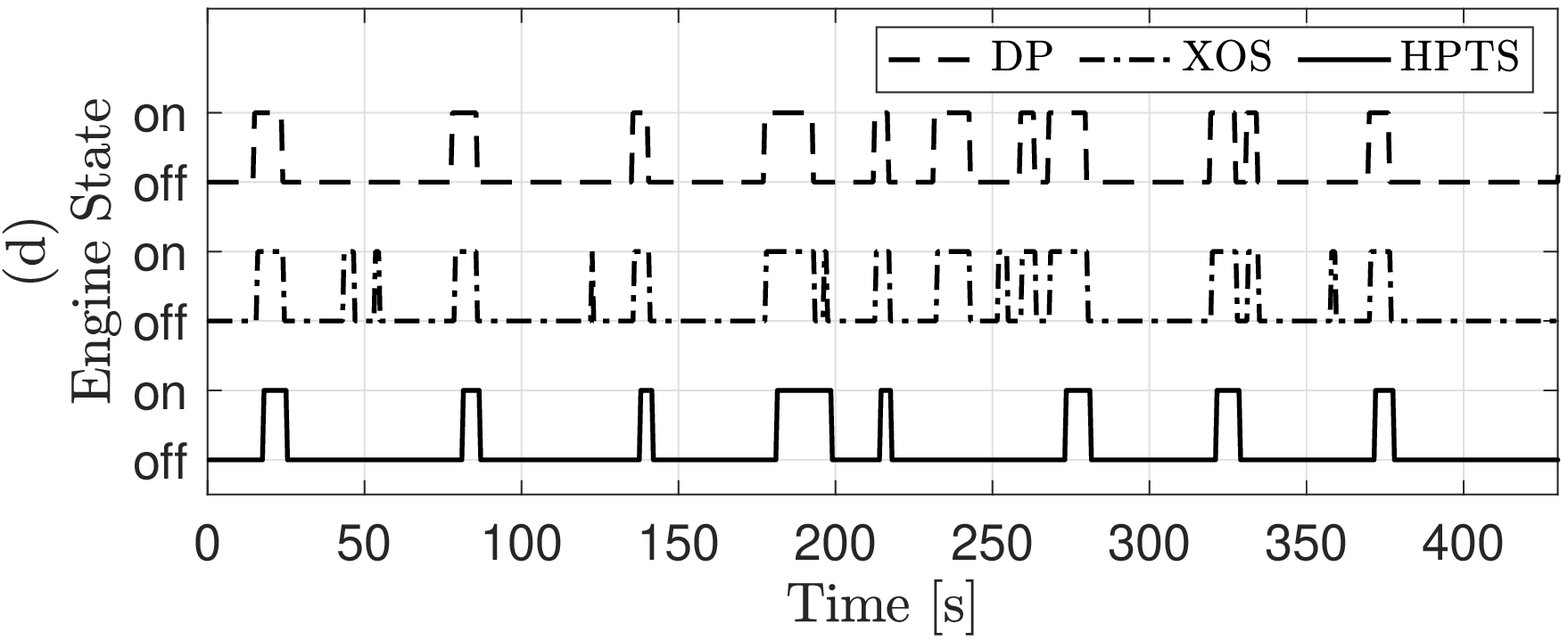}\\[-2ex]
\caption{{Power ((a): DP, (b): XOS, (c): HPTS) 
and engine state $s(t)$ (plot (d)) profiles when WL-M is simulated with the linear FCM model \eqref{eq:Q_f} and CS condition $\Delta \soc\!=\!0$. 
}}
\label{fig:powerprofile_linear}
\end{figure}

Fig.~\ref{fig:powerprofile_linear} presents the power profiles {and the associated engine on/off states} determined by these control methods when the WL-M cycle is simulated. 
\begin{figure}[htb!]
\centering
\includegraphics[width=\columnwidth]{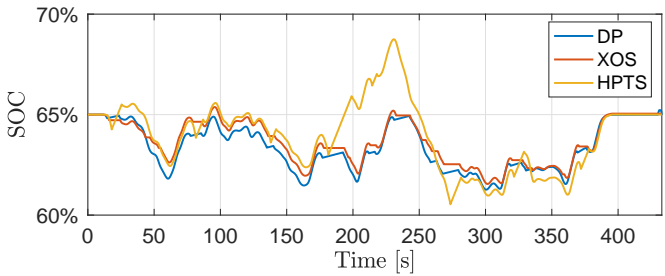}\\[-2.1ex]
\caption{Battery SOC profiles when WL-M is simulated with the linear FCM model \eqref{eq:Q_f}. CS condition $\Delta \soc\!=\!0$ is achieved in all cases.}
\label{fig:socprofile_linear}
\end{figure}
Compared to the XOS, both DP and the {\emstwo} can reduce the engine restarts by manipulating the operation of the PS and SS. In particular, when the ICE is activated in cases of DP and {\emstwo}, the powertrain is operated in hybrid mode, in which the battery is charged by the ICE at a constant SS power, and therefore additional engine breaks are allowed  to prevent the engine status from being changed too frequently, as can be found in the control solution of XOS (for example, around 50\,s and 250\,s in Fig.~\ref{fig:powerprofile_linear}). 
Further comparing the solutions of {\emstwo} and DP, the {\emstwo} is more sluggish in its response to a PS power request (a phase delay can be observed in Fig.~\ref{fig:powerprofile_linear} by comparing their PS profiles) due to the impact of the hysteresis switching mechanism. Moreover, the PS operating power of {\emstwo} is higher than DP, which also yields more battery charge during $s=1$ and further reduced ICE on/off transitions. As a consequence, the battery SOC of {\emstwo} has more noticeable variations than the profiles generated by the other two methods, as shown in Fig. \ref{fig:socprofile_linear}. For example, the battery in the case of {\emstwo} is charged intensively from 180s to 220s, thereby allowing the ICE to be switched off for the next 50s. Although the SOC in the case of XOS performs closely to the DP solution, it will be shown later that the fuel economy of XOS is significantly impaired by the unnecessary ICE switches (which incur additional fuel usage). The fuel consumption of all the methods is compared in Table~\ref{table:fuelconsumption_linear}.

\begin{table}[htb!]
\centering
\caption{Fuel consumption $[g]$ with the linear FCM model and percentage fuel increase compared to DP solutions.}
\label{table:fuelconsumption_linear}
    \begin{tabular*}{0.95\columnwidth}{l @{\extracolsep{\fill}} cccc}
    \hline
    \hline
       &  DP &  XOS  & {\emstwo}\\
    \hline
    WL-L         & 54.5  & 61.3 (12.5\%) & 57.8 (6.06\%) \\ 
   WL-M          & 99.5  & 106.5 (7.04\%)& 104.2 (4.72\%) \\
   WL-H          & 175.1 & 183.3 (4.68\%)& 179.8 (2.68\%) \\
    WL-E         & 311.1 & 313.4 (0.74\%)& 312.6 (0.48\%) \\
    Experimental &251.2  & 273.6 (8.92\%) & 261.1 (3.49\%) \\
    \hline
    \hline
    \end{tabular*}
\end{table}

The solutions of the {\emstwo} are much closer to the results of DP as compared to the XOS for all cases. Approximately, there is 0.48\%-6.06\% more fuel usage by {\emstwo} than the DP for WLTP cycles, and the gap decreases from WL-L to WL-E. This can be understood that more fuel saving is expected by an optimally controlled SSS during urban driving rather than driving on the motorway, where the SSS is rarely engaged.
In the context of the real-world experimental driving cycle that involves mixed traffic conditions, the {\emstwo} solution is only 3.49\% behind the DP and can save about 5.4\% more fuel as compared to the XOS. The results verify the capability of {\emstwo} in dealing with general driving scenarios.

\begin{figure}[htb!]
\centering
\includegraphics[width=\columnwidth]{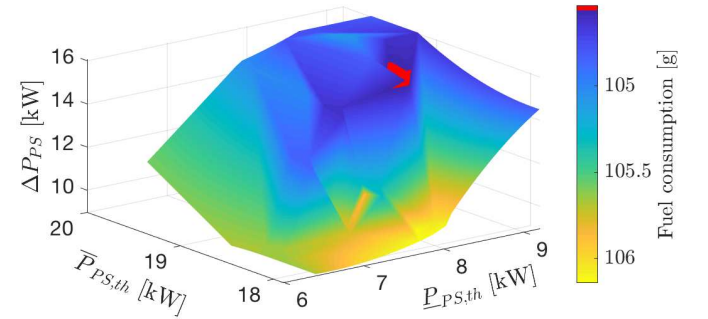}\\[-2ex]
\caption{HPTS optimal solutions obtained by tuning $\overline{P}_{PS,th}$, $\underline{P}_{PS,th}$ and $\Delta P_{PS}$, when the nonlinear experimental FCM(see Fig.~\ref{fig:fuelmassrate}) is employed and while satisfying charge sustaining operation.}
\label{fig:HPFS_search_nonlinear}
\end{figure}

\begin{figure}[htb!]
\centering
\includegraphics[width=\columnwidth]{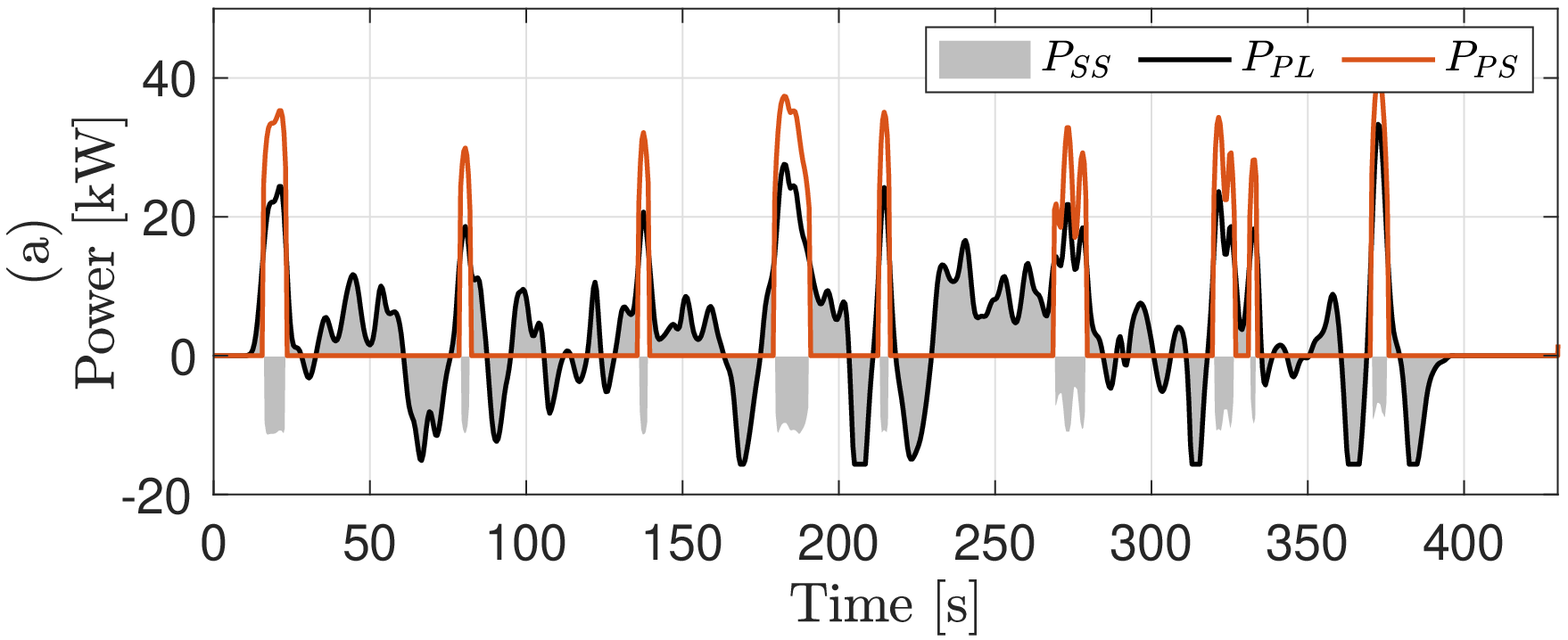}\\[-4.3ex]
\includegraphics[width=\columnwidth]{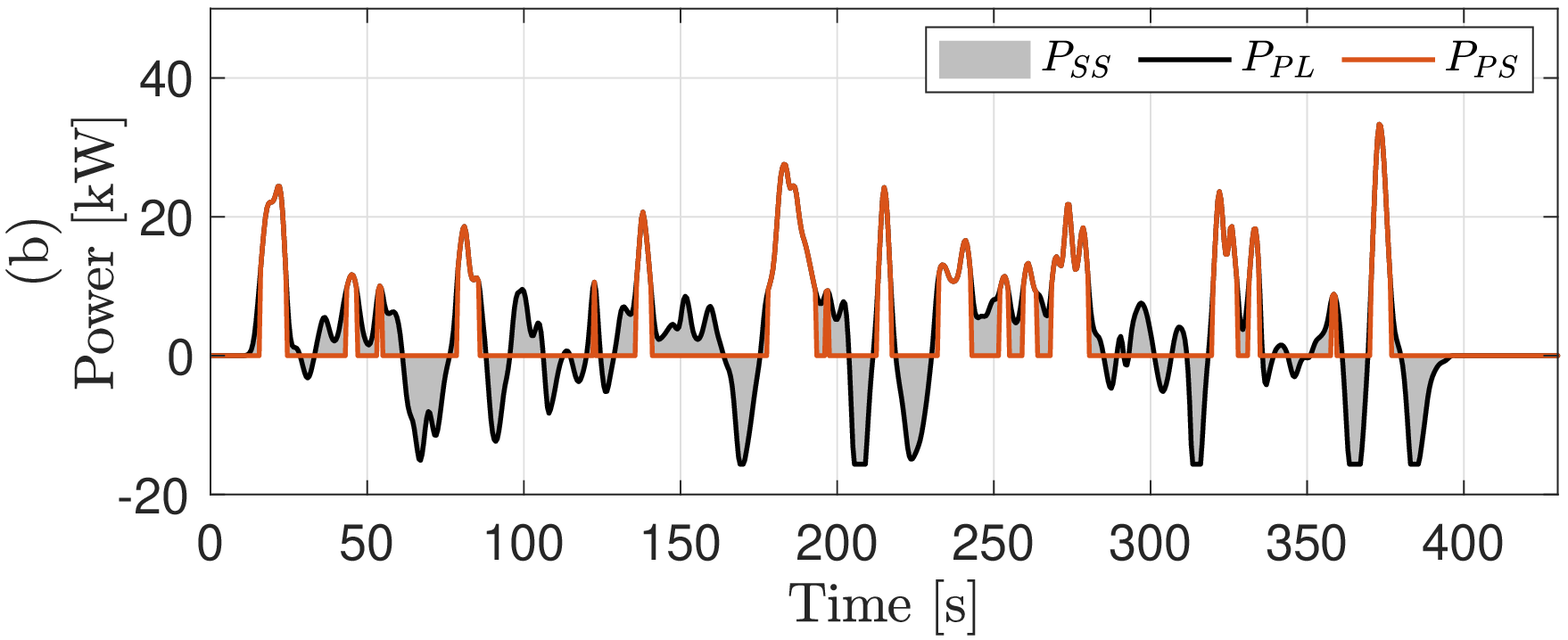}\\[-4.3ex]
\includegraphics[width=\columnwidth]{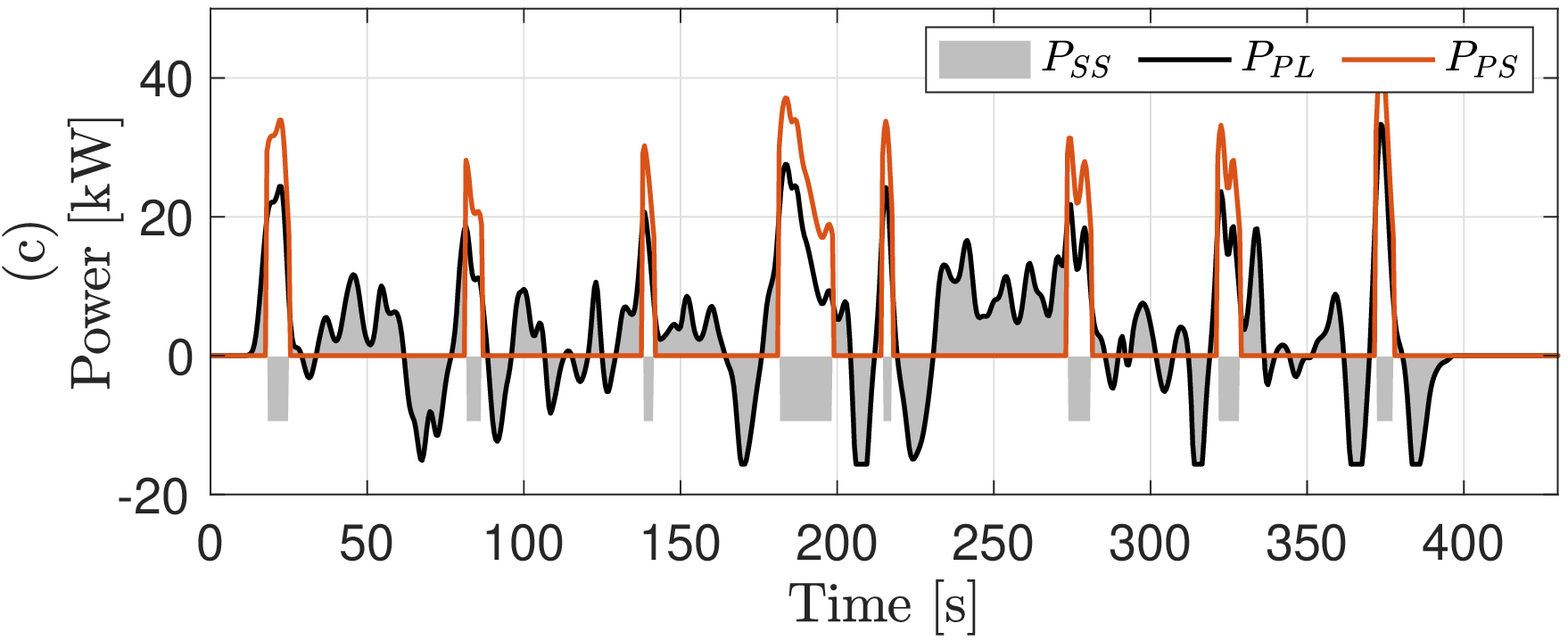}\\[-4.3ex]
\includegraphics[width=\columnwidth]{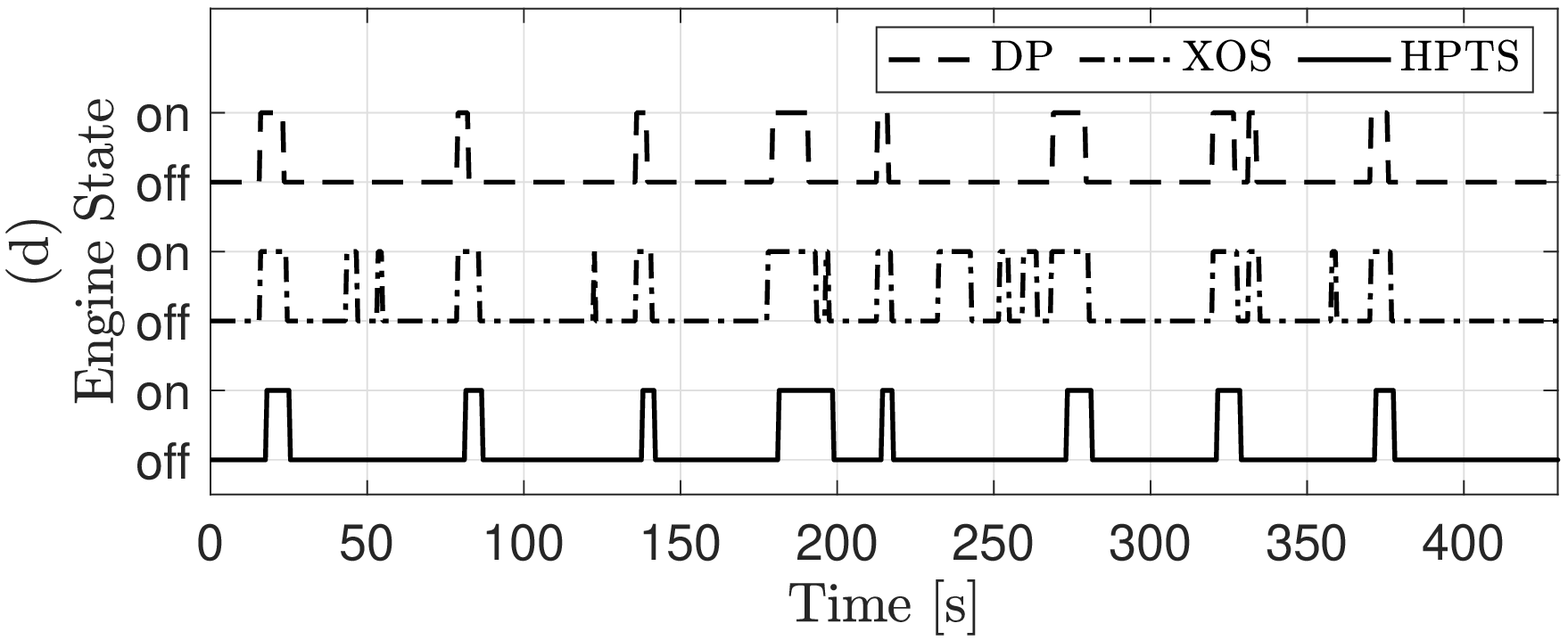}\\[-2ex]
\caption{{Power profiles and engine switching profile when WL-M is simulated with the nonlinear experimental FCM and CS condition $\Delta \soc\!=\!0$. (a): DP. (b): XOS. (c): HPTS. (d): engine state $s(t)$.}}
\label{fig:powerprofile_nonlinear}
\end{figure}

\begin{figure}[htb!]
\centering
\includegraphics[width=\columnwidth]{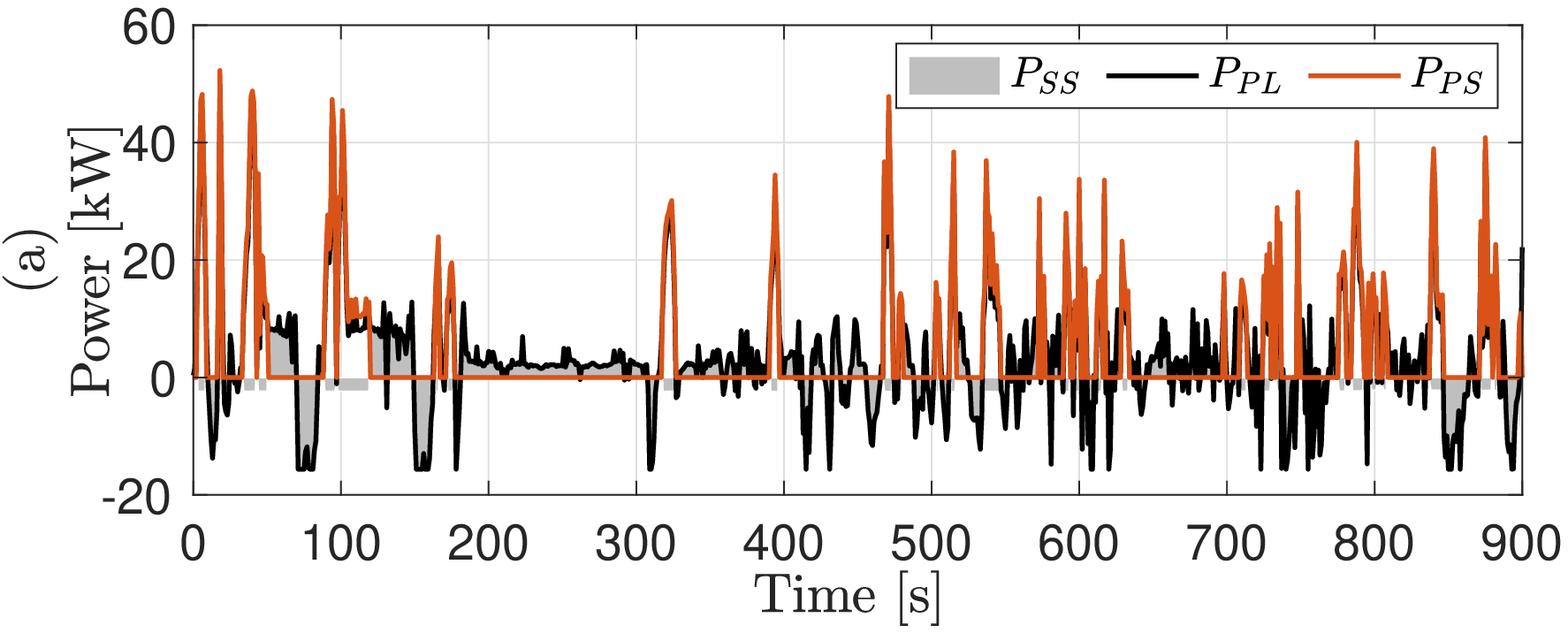}\\[-4.3ex]
\includegraphics[width=\columnwidth]{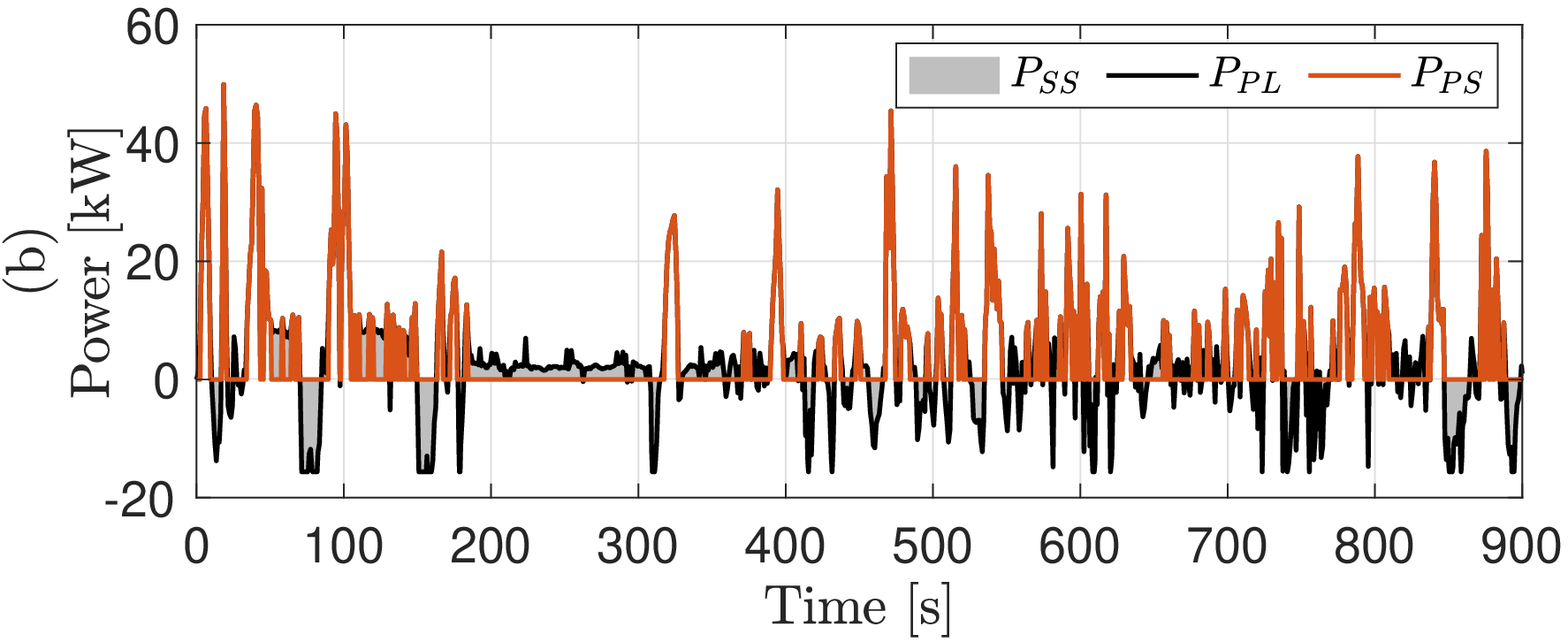}\\[-4.3ex]
\includegraphics[width=\columnwidth]{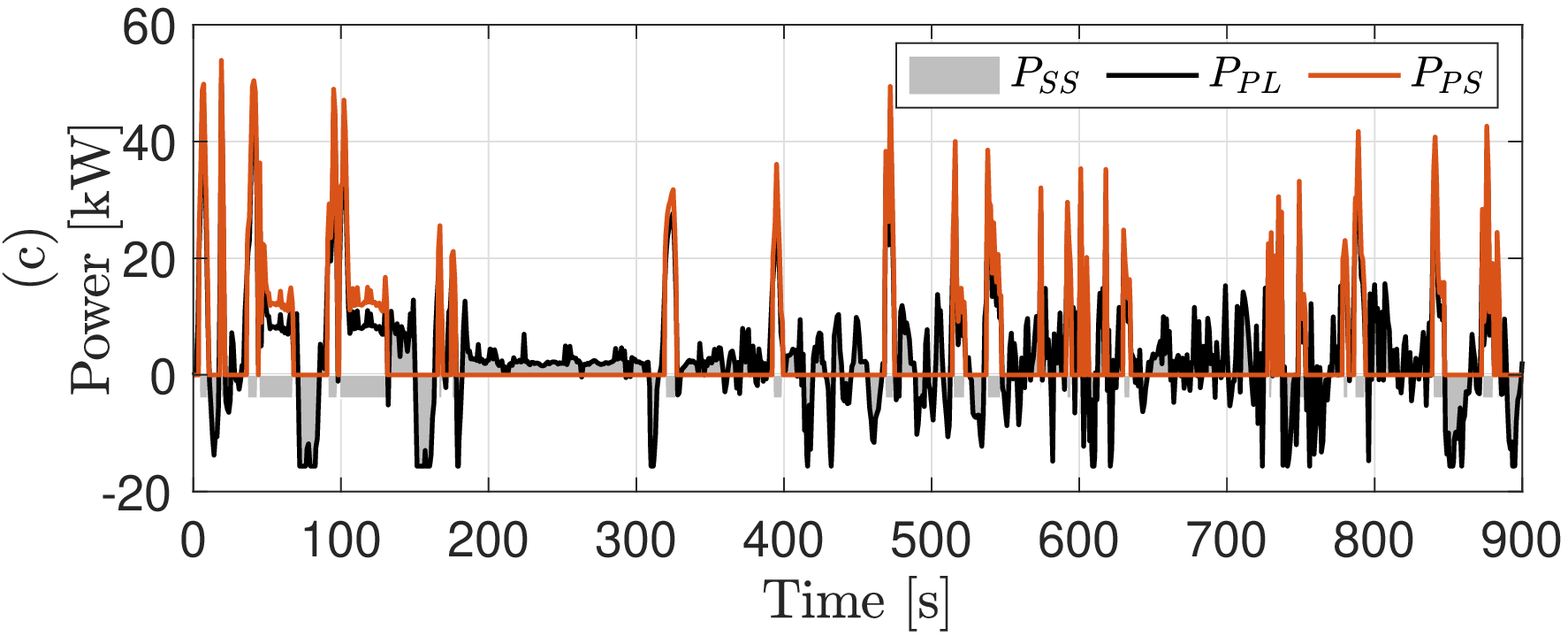}\\[-4.3ex]
\includegraphics[width=\columnwidth]{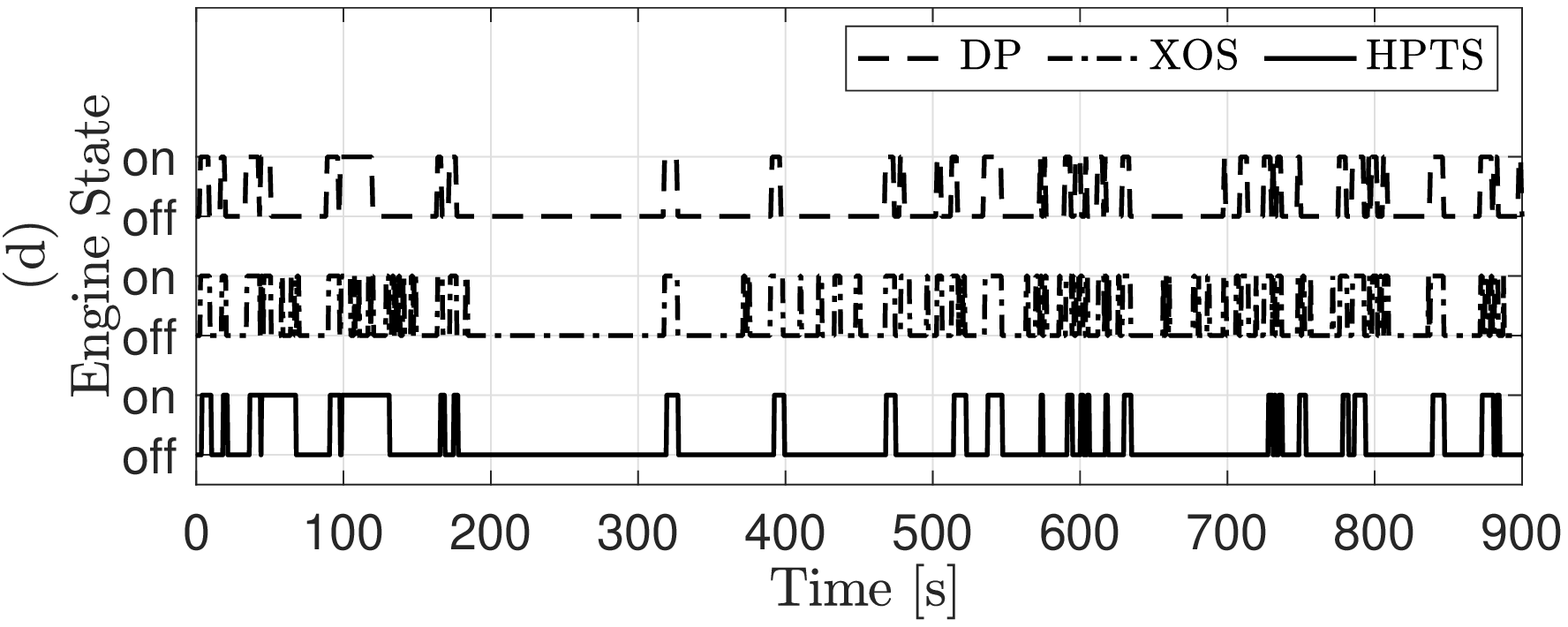}\\[-2ex]
\caption{{Power profiles and engine switching profile when the experimental driving cycle is simulated with the nonlinear experimental FCM and CS condition $\Delta \soc\!=\!0$. (a): DP. (b): XOS. (c): HPTS. (d): engine state $s(t)$.}}
\label{fig:powerprofile_ex}
\end{figure}

The performance of the {\emstwo} is further examined by employing the experimental FCM shown in Fig.~\ref{fig:fuelmassrate}.  
Similarly to the linear FCM case, the optimal selection of the design parameters is identified by a global tuning process. As shown in Fig.~\ref{fig:HPFS_search_nonlinear}, the optimal parameter selection for the nonlinear FCM is $\overline{P}_{PS,th}\!=\!18.75$\,kW, $\underline{P}_{PS,th}\!=\!8.5$\,kW and $\Delta P_{PS}\!=\!15.0$\,kW.%
The power {and engine switching} profiles of the three control methods are shown in Fig.~\ref{fig:powerprofile_nonlinear}. 
\begin{figure}[htb!]
\centering
\includegraphics[width=\columnwidth]{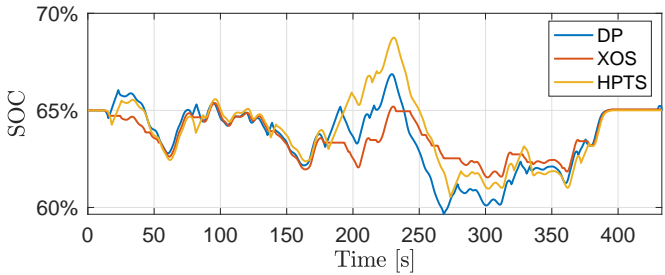}\\[-2.1ex]
\caption{Battery SOC profiles when WL-M is simulated with the nonlinear experimental FCM. CS condition $\Delta \soc\!=\!0$ is achieved in all cases.}
\label{fig:socprofile_nonlinear}
\end{figure}
As it can be seen, the {\emstwo} is able to emulate the power profiles solved by DP while XOS remains at the same operation as shown in Fig.~\ref{fig:powerprofile_linear}. {The engine switching profiles further confirm the findings; it is notable that once again both DP and HPTS reduce significantly the engine on/off events compared to the XOS.} The SOC profiles reported in Fig.~\ref{fig:socprofile_nonlinear} show that, in this case, the {\emstwo} can produce a profile closer to the DP solution as compared to the XOS. {To provide further evidence of the resemblance between DP and the {\emstwo}, the control solutions of the three methods for the experimental driving cycle are also demonstated in Fig.~\ref{fig:powerprofile_ex}. As it can be seen, both the DP and {\emstwo} have similar power profiles while the XOS entails much more engine switching operations, which incur additional fuel usage.} The fuel consumption results of all the methods in the case of experimental FCM are presented in Table~\ref{table:fuelconsumption_nonlinear}.
\begin{table}[ht!]
\centering
\caption{Fuel consumption $[g]$ with the nonlinear experimental FCM and percentage fuel increase compared to DP solutions.}
\label{table:fuelconsumption_nonlinear}
    \begin{tabular*}{0.95\columnwidth}{l @{\extracolsep{\fill}} cccc}
    \hline
    \hline
       &  DP &  XOS  & {\emstwo}\\
    \hline
    WL-L         & 55.0  &  68.1 (23.8\%) & 59.0 (7.27\%)  \\ 
   WL-M          & 99.7  &  113 (13.3\%) & 105.8 (6.12\%) \\
   WL-H          & 174.8 &  192.1 (9.90\%)& 181.0 (3.55\%) \\
    WL-E         & 307.1 &  312.3 (1.69\%)& 309.8 (0.88\%) \\
    Experimental & 249.4  & 286.1 (14.72\%)  & 259.9 (4.21\%)  \\
    \hline
    \hline
    \end{tabular*}
\end{table}
By comparing the results with the previous solutions solved for the linear FCM model in Table~\ref{table:fuelconsumption_linear}, it can be observed that the optimality (percentage fuel increase) of the XOS degrades considerably, while the {\emstwo} is more robust against the model nonlinearity, with the optimality decreased by only 0.4\%-1.4\% for each cycle as compared to the linear case.
It is noteworthy that for the experimental driving cycle the fuel increase for the HPTS is only 0.04\% as compared to its counterpart with the linear FCM, and furthermore the HPTS achieves an astonishing 10\% less fuel consumption as compared to the XOS. 

\begin{figure}[htb!]
\centering
\includegraphics[width=\columnwidth]{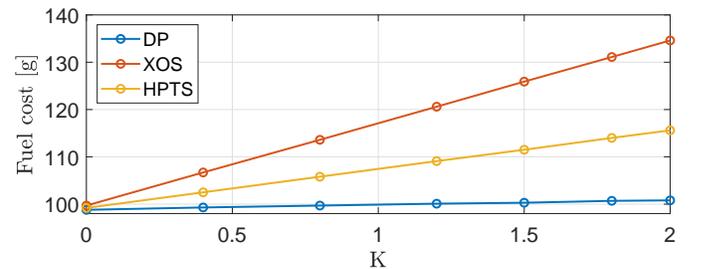}\\[-2.1ex]
\caption{Fuel consumption cost with varied penalty fuel coefficient $K$ using the nonlinear experimental FCM and CS condition $\Delta \soc\!=\!0$ when WL-M is simulated.}
\label{fig:powerprofile_nonlinear_K}
\end{figure}

To gain more insight into the effect of the penalty fuel for the engine reactivation, a further investigation is carried out to compare the solutions of the three control methods using different penalty fuel coefficient $K$. For $K \in \left[0,2\right]$, the total fuel consumption of the three methods for the experimental FCM is depicted in Fig.~\ref{fig:powerprofile_nonlinear_K}.
As it can be noticed, the {\emstwo} is found to outperform the XOS for all studied $K$ with up to 16.44\% improvement in terms of fuel consumption. Moreover, the fuel usage is linearly dependent on the penalty fuel coefficient $K$ for all the three control cases, and the differences between each solution of the tested methods decrease as $K$ decreases, which is expected with the further improvement of the SSS efficiency. When $K=0$, the {\emstwo} highly resembles the global optimal solution delivered by DP, with only $0.4\%$ fuel increment, whereas the XOS lags the {\emstwo} by another $0.5\%$.

The global tuning (by repetitive simulation for a batch of parameter combinations) of HPTS and the running time of DP required for the solutions reported in Table~\ref{table:fuelconsumption_nonlinear} are further compared in Table~\ref{table:comparison}. The evaluation of both methods is performed in Matlab \& Simulink environment on an Intel i5 2.9 GHz CPU with 8 GB of memory. As it can be seen, the proposed HPTS is more computationally efficient than DP, while the benefit is expected to become more significant for a more complex powertrain model as the computational burden of DP increases exponentially with the number of system states (eventually DP becomes unusable even for moderately complex models). It is not difficult to see that the tuning effort depends on the size of the power interval of searching, which is usually identified empirically. With the tuning results obtained for more tested cycles, more accurate searching intervals can be identified when a new driving cycle is investigated based on the nature of the cycle (for example, urban, rural or motorway driving), and therefore the tuning effort can be further reduced. Moreover, HPTS acts entirely on the three tunable control parameters, as opposed to classic optimal control (DP) that acts on a state input. This is another salient feature that allows the globally tuned HPTS parameters for any test cycles (pre-determined and available in the database, such as WLTP) to be directly applied to any unknown cycle based on the speed forecast, such as the predicted average speed from a navigation system and classification against the test cycles. In such a case, the tuning effort is negligible. 

\begin{table}[H]
\centering
\caption{Comparison of tuning (HPTS) and running time (DP) [minutes] required by HPTS and DP for the results in Table~\ref{table:fuelconsumption_nonlinear}.}
\label{table:comparison}
    \begin{tabular*}{0.99\columnwidth}{l @{\extracolsep{\fill}} cccccc}
    \hline
    \hline
   		       & WL-L & WL-M & WL-H & WL-E & Experimental   \\  \hline
   DP			& 61.81 & 43.89 & 47.44 & 34.12 & 93.02   \\
   HPTS			& 17.09 & 15.50 & 15.70 & 14.77 & 18.58		\\
    \hline
    \hline
    \end{tabular*}
\end{table}
\section{Conclusions}\label{sec:conclusions}
This paper proposes a novel rule-based energy management (EM) control strategy for series hybrid electric vehicles (HEVs) with the engine start-stop system (SSS), the hysteresis power threshold strategy ({\emstwo}). The principal mechanisms of the {\emstwo} are developed with inspiration from the closed-form solutions of the optimal energy source power split derived in this paper.
In particular, the mathematical analysis is carried out for two model cases: 1) without SSS, 2) with the lossless SSS where the fuel usage for engine restarts is ignored, thus yielding two fundamental optimization solutions that can be represented by simple control rules. The {\emstwo} further extends these rules with consideration of a more realistic SSS model that incorporates penalty fuel for engine restarts. The {\emstwo} essentially combines two operational modes: battery-only mode and hybrid/engine-only mode, with the latter mode depending on a tuneable power offset. The two modes are separated by a hysteresis switching algorithm parameterized by a pair of thresholds. As such, a minimum duration is ensured for each mode, which naturally prevents fast engine on/off switching that is detrimental to fuel usage. 
The two thresholds and the offset are regulated based on the information of different HEV model (or real vehicle) parameters and driving cycles by a systematic tuning process, targeting charge sustaining operation that is proven in this paper to be optimal in the context of the equivalent fuel consumption. 

DP simulation results verify the analytic solutions obtained for the two simple vehicle models. As such, the globally optimal solutions for these models can be simply produced without referring to dynamic programming (DP), which usually involves heavy computation effort. The control performance of the {\emstwo} is evaluated and benchmarked against DP and a recently proposed rule-based method, the exclusive operation strategy (XOS), in simulations with a realistic SSS that involves a fuel penalty for engine switch on. 
It is demonstrated that the proposed {\emstwo} outperforms the XOS for all studied driving cycles, especially for the profiles that emulate urban driving. Moreover, the simple nature of the {\emstwo} also makes it a potential benchmarking strategy for high-fidelity vehicle models, where DP is no longer applicable. 
Future work involves extending {\emstwo} to incorporate driving speed prediction, which could allow the {\emstwo} to be implemented in real-time. 


\bibliographystyle{IEEEtran}
\bibliography{ref}
\end{document}